\documentclass[twocolumn,prb,superscriptaddress]{revtex4-1}
 
\usepackage{graphicx,amsmath,xcolor,soul}
\usepackage[capitalise]{cleveref} 
\usepackage{xcolor} 
\usepackage{amsfonts}

\usepackage{titlesec}
\titleformat{\section}{\normalfont\large\bfseries\raggedright}{\thesection}{1em}{}
\titleformat{\subsection}{\normalfont\bfseries\raggedright}{\thesubsection}{1em}{}
\titlespacing*{\section}
{0pt}{0ex}{0ex}
\titlespacing*{\subsection}
{0pt}{0ex}{0ex}

\begin{document}

\title{Topological polarons in halide perovskites}

\author{Jon Lafuente-Bartolome}
\author{Chao Lian}
\affiliation{Oden Institute for Computational Engineering and Sciences, The University of Texas at Austin, Austin, Texas 78712, USA}
\affiliation{Department of Physics, The University of Texas at Austin, Austin, Texas 78712, USA}
\author{Feliciano Giustino}
\email{fgiustino@oden.utexas.edu}
\affiliation{Oden Institute for Computational Engineering and Sciences, The University of Texas at Austin, Austin, Texas 78712, USA}
\affiliation{Department of Physics, The University of Texas at Austin, Austin, Texas 78712, USA}

\date{\today}

\begin{abstract}
Halide perovskites emerged as a revolutionary family of high-quality semiconductors for solar energy harvesting and energy-efficient lighting. There is mounting evidence that the exceptional optoelectronic properties of these materials could stem from unconventional electron-phonon couplings, and it has been suggested that the formation of polarons and self-trapped excitons could be key to understanding such properties. By performing first-principles simulations with unprecedented detail across the length scales, here we show that halide perovskites harbor a uniquely rich variety of polaronic species, including small polarons, large polarons, and charge density waves, and we explain a variety of experimental observations. We find that these emergent quasiparticles support topologically nontrivial phonon fields with quantized topological charge, making them the first non-magnetic analog of the helical Bloch points found in magnetic skyrmion lattices.
\end{abstract}

\maketitle

Halide perovskites have rapidly emerged as a unique class of materials with outstanding optoelectronic properties \cite{Green2014,Stranks2015,Huang2017,Stranks2013,Xing2013,Brenner2016}, and are being considered for applications in photovoltaics \cite{Kojima2009,Lee2012,Yang2015,Min2021}, X-ray detection \cite{Wei2016}, solid-state lighting \cite{Liu2021}, lasers \cite{Zhu2015,Sutherland2016}, and photocatalysis \cite{Park2016}. These developments have triggered a race to uncover the fundamental microscopic processes that govern the photophysics of these materials \cite{Huang2017}. Several novel phenomena have been identified, for example crystal-liquid duality \cite{Zhu2016,Miyata2017}, dynamic Rashba effects \cite{Niesner2018,Schlipf2021}, and hot-phonon bottlenecks \cite{Price2015,Yang2016}.
The common thread across these diverse phenomena is an unconventional coupling between electrons and phonons \cite{Wright2016}.

It is believed that the electron-phonon interaction in halide perovskites plays a crucial role in their photophysical properties by promoting the formation of polarons \cite{Miyata2018,Zhu2015_2, Zhu2016, Qian2023,Luo2018,Wu2021,ZhangH2023}. Polarons are localized electron or holes accompanied by a cloud of phonons which distorts the surrounding ionic lattice \cite{Franchini2021,Buizza2021}. It has been suggested that large ferroelectric polarons \cite{Miyata2018} might protect charge carriers against scattering and recombination in lead halide perovskites \cite{Zhu2015_2, Zhu2016, Qian2023}, and small polarons have been invoked to explain efficient luminescence from halide double perovskites \cite{Luo2018,Wu2021}. Polaron dynamics was recently observed via ultrafast experiments in both three-dimensional and two-dimensional halide perovskites \cite{Guzelturk2021,ZhangH2023}. 

Despite growing evidence of the role of polarons in halide perovskites, the nature of these quasiparticles remains poorly understood. Assignments from experimental studies range from intermediate-coupling Fr\"ohlich polarons \cite{Miyata2017_2,Batignani2018,Puppin2020,Guzelturk2021} to strong-coupling acoustic polarons \cite{Wu2021}, as well as Jahn-Teller polarons \cite{Luo2018}. Theoretical attempts at reconciling these experimental observations have met with limited success, because atomic-scale density-functional theory calculations are restricted to small polarons \cite{Neukirch2016,Osterbacka2020} and periodic superstructures \cite{Meggiolaro2020,Cannelli2021, Qian2023}.

Here, we provide a unified picture of polarons in halide perovskites by deploying a computational approach that allows us to probe these quasiparticles from the nanoscale to the mesoscale and for varying carrier concentrations \cite{Sio2019,Sio2019_b,Lafuente2022,Lafuente2022_b}. Our data indicate that halide perovskite host three distinct polaron species, namely large polaronic helical Bloch points, small ferrotoroidic polarons, and periodic twist-density waves. Two of these species are found to possess topological properties.

\medskip

\section*{Results}

\subsection*{Large polaronic helical Bloch points}

\begin{figure*}
\centering
\includegraphics[width=0.9\textwidth]{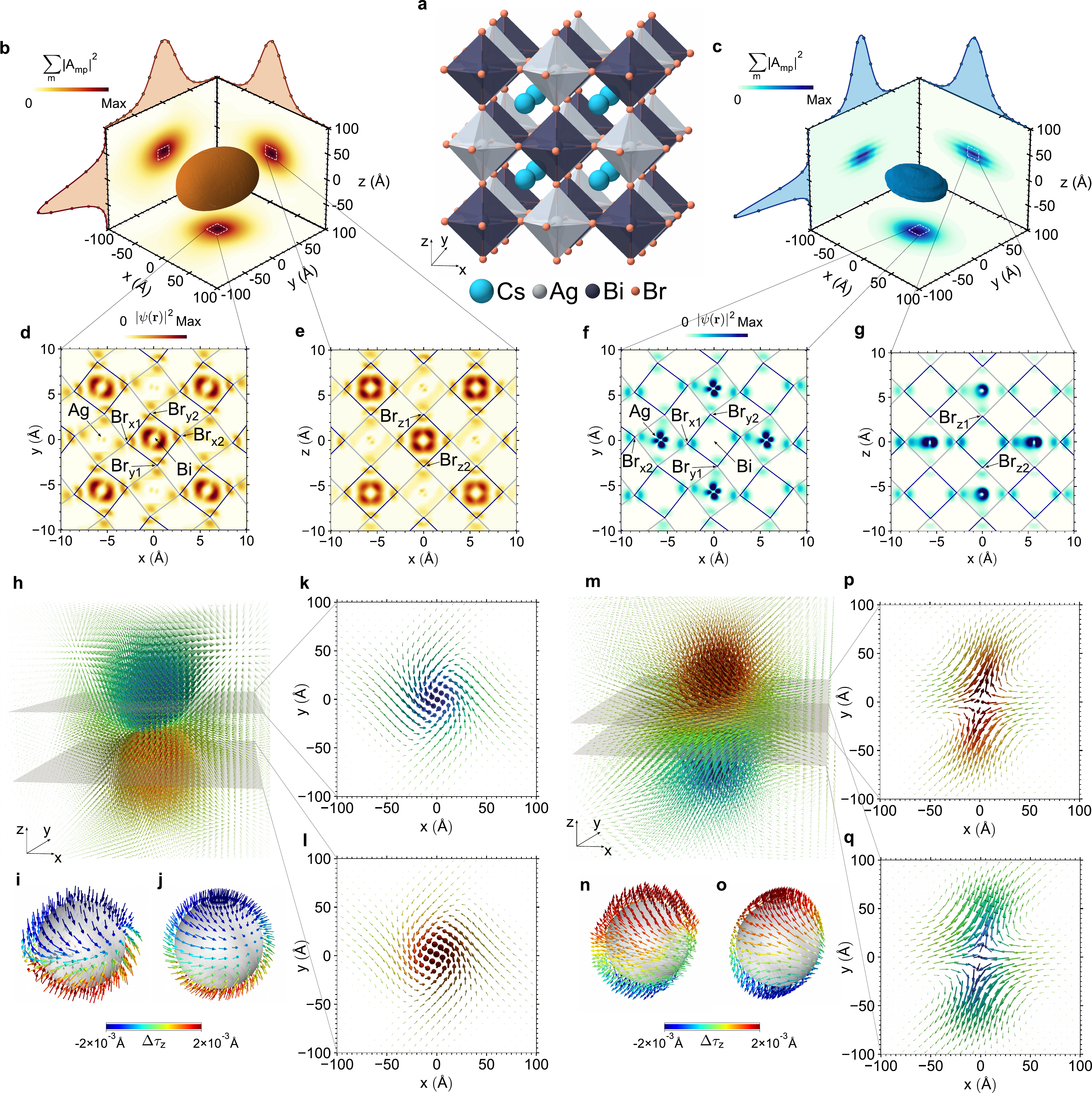}
\caption{\textbf{Large helical Bloch point polarons in halide perovskites.} \textbf{a} Structural model of Cs$_2$AgBiBr$_6$ with $I4/m$ space group. The Cartesian axes $x$, $y$, $z$, are aligned with the orthogonal axes of the pseudocubic unit cell (cf. Supplementary Note 3). \textbf{b} Envelope of the wavefunction of electron polaron in a non-diagonal $20 \times 20 \times 20$ pseudocubic supercell of Cs$_2$AgBiBr$_6$ (cf. Supplementary Note~1 and 3), as obtained from the square moduli of the Wannier function coefficients, $|A_m({\bf R}_p)|^2$, defined in Eq.~(S10) of Supplementary Note~2. The isovalue is set to $10\%$ of the maximum. Cross-sections through the polaron center are shown on the axis planes, and linear profiles through the polaron center are shown above each axis. A spline interpolation is performed between discrete values at the lattice sites, shown as discs. The standard deviations of these profiles are $\sigma_{x}=24$~\AA, $\sigma_y=29$~\AA, $\sigma_z=27$~\AA. \textbf{c} Same as in (b), but for the hole polaron. The standard deviations of the line profiles are $\sigma_{x}=23$~\AA, $\sigma_y=17$~\AA, $\sigma_z=13$~\AA. \textbf{d}, \textbf{e}
Close-up views of the square modulus of the wavefunction of the electron polaron along an $xy$-plane and an $xz$-plane through the polaron center, respectively. \textbf{f}, \textbf{g} Same as (d), (e) for the hole polaron. \textbf{h} 3D view of the atomic displacements associated with the large electron polaron in Cs$_2$AgBiBr$_6$. For clarity only Ag displacements are shown. The color code represents the magnitude of the displacement in the $z$ direction. \textbf{i}, Displacement patters from (h), visualized on a sphere enclosing the polaron center. Displacements have been normalized for clarity. This vector field corresponds to a helical Bloch point with topological charge $Q=-1$, vorticity $q=1$, polarity $p=-1$ and helicity $\gamma=3\pi/4$ (cf. Supplementary Fig.~S4). \textbf{j} Ideal Bloch point with the same topological charge, vorticity, polarity and helicity, for comparison to the electron polaron in (i). \textbf{k}, \textbf{l} Top view of the Ag displacements from (h), along the cross-sections shown in (h). These vector fields have the same vortex patterns of magnetic merons. In fact, surface integrals of the Berry curvature on each plane yield a 2D topological charge $0.39<|Q|<0.43$, which is close to that of merons, $|Q|=0.5$ (cf.\ Fig.~S4). \textbf{m}, \textbf{n}, \textbf{o}, \textbf{p}, \textbf{q} Same as (h), (j), (i), (k), (l) but for the hole polaron. In this case the Ag displacements define a vector field corresponding to a Bloch point with topological charge $Q=-1$, vorticity $q=-1$, polarity $p=1$ and helicity $\gamma=0.9\pi$, and whose plane cuts exibit antivortex patterns as in magnetic merons (cf.\ Fig.~S5).}
\label{fig1}
\end{figure*}

We focus on the lead-free double perovskite Cs$_2$AgBiBr$_6$ as a representative member of the halide perovskite family with complex electron-phonon interactions \cite{Wu2021,Wright2021,He2023} and without the added complication of organic cations. Figure~\ref{fig1}\textbf{a} shows the low-temperature tetragonal structure \cite{Schade2019}, which consists of corner-sharing AgBr$_6$ and BiBr$_6$ octahedra alternating in a rock-salt lattice (cf.\ Supplementary Note~1 for computational details). Figures~\ref{fig1}\textbf{b} and \textbf{c} show the envelopes of the electron wavefunction and the hole wavefunction, which describe the spatial distribution of an extra electron added or removed from the system, respectively (cf.\ Supplementary Note~2). These envelopes follow smooth gaussian profiles, which are reminiscent of the textbook Fr\"ohlich polaron model \cite{Pekar1946,Feynman1955}. The electron polaron is nearly spherical in shape, has a full-width at half maximum of $6.3\pm 0.6$~nm, and spans eight crystal unit cells. Therefore it can be described as a large polaron. The hole polaron is slightly more compact with a full-width at half maximum of $4.2\pm 1.2$~nm, and resembles an oblate spheroid with the short axis along the apical direction. This anisotropy is inherited from the effective mass tensor of the valence bands \cite{Leveillee2021}. 

Zooming in at the scale of individual octahedra, Figs.~\ref{fig1}\textbf{d} and \ref{fig1}\textbf{e} show that the large electron polaron draws weight from Bi-$6p$ and Br-$4p$ orbitals; while the large hole polaron consists primarily of Ag-$4d$ and Br-$4p$ states (Figs.~\ref{fig1}\textbf{f} and \ref{fig1}\textbf{g}). These compositions reflect the characters of the conduction and valence band edges, as shown in Fig.~S1. 

So far, the structure of large polarons in halide perovskites appears similar to that of simple ionic compounds such as the alkali halides \cite{Sio2019,Lafuente2022}. However, inspection of the phonon component of these polarons reveals unexpected features: Fig.~\ref{fig1}\textbf{h} shows that the atomic displacements in the large electron polaron follow a swirling pattern around the polaron center. Two-dimensional cross-sections of this 3D object in Figs.~\ref{fig1}\textbf{k} and \ref{fig1}\textbf{l} reveal a well-defined vortex structure. A similar pattern is seen for the large hole polaron in Fig.~\ref{fig1}\textbf{m}, except that in this case the 2D cross-sections (Figs.~\ref{fig1}\textbf{p} and \ref{fig1}\textbf{q}) reveal an antivortex. An enlarged view of the displacement patterns for the large electron and hole polarons are shown in Supplementary Figs.~S2 and S3, respectively.

These displacements patterns carry a strong resemblance to the spin textures of magnetic skyrmions \cite{Nagaosa2013,Fert2017}, suggesting that polarons in halide perovskites might possess topological properties. To investigate this possibility, in Figs.~\ref{fig1}\textbf{i} and \ref{fig1}\textbf{n} we visualize the vector field of displacements on a sphere enclosing the polaron center. For both polaron species, this field wraps the sphere, and gradually evolves from tangential character at the equator to radial character at the poles. From the topological classification adopted in the skyrmion literature, we conclude that these textures are closest to the helical Bloch points found in magnetic systems \cite{Doring1968,Gobel2021} and shown in Figs.~\ref{fig1}\textbf{j}, \ref{fig1}\textbf{o}, and Fig.~S4. To place this assignment on quantitative grounds, we evaluate the quantized topological charge $Q$ as the flux of the Berry curvature $\boldsymbol{\Omega}$ through the unit sphere: $Q = (1/4\pi) \int \boldsymbol{\Omega}\cdot d{\bf S}$, with $\boldsymbol{\Omega}_\alpha = \varepsilon_{\alpha\beta\gamma} \mathbf{u} \cdot ( \partial \mathbf{u}/\partial x_\beta \times \partial \mathbf{u}/\partial x_\gamma )$. Here, ${\bf u}(x_1,x_2,x_3)$ is the atomic displacement field, Greek subscripts denote Cartesian directions, $\varepsilon_{\alpha\beta\gamma}$ is the Levi-Civita symbol, and $d{\bf S}$ is the surface element. The charge $Q$ counts how many times the displacement field wraps around the unit sphere \cite{Nagaosa2013,Yu2018}. For 2D topological structures, this integral is evaluated over a plane: skyrmions correspond to $Q=\pm 1$ \cite{Fert2017}, and merons correspond to $Q=\pm 1/2$ \cite{Yu2018}. For 3D structures, the topological charge is $Q=\pm 1$ for Bloch points \cite{Hierro2020}. Our calculations indicate that $Q = 1$ for both the electron and the hole polaron, confirming that we are in the presence of topological quasiparticles with quantized topological charge. Further analysis shown in Figs.~S4 and S5 indicates that the electron and hole polarons are helical Bloch points, with vorticities $q=1$, $q=-1$ and helicities $\gamma=3\pi/4$, $\gamma=0.9\pi$ for the electron and hole polarons, respectively. In addition, the cross-sectional views in Figs.~\ref{fig1}\textbf{k}, \ref{fig1}\textbf{l}, and \ref{fig1}\textbf{p}, \ref{fig1}\textbf{q} carry the topology of magnetic merons and antimerons, respectively. The present discovery of topological polarons raises the tantalizing prospect that these quasiparticles may also exhibit emergent phenomena such as topological and skyrmion Hall effects through the anomalous magnetic field associated with their Berry curvature~\cite{Nagaosa2013}.

\medskip

\subsection*{Small ferrotoroidic polarons}

\begin{figure*}[hbt!]
    \centering
    \includegraphics[width=0.95\textwidth]{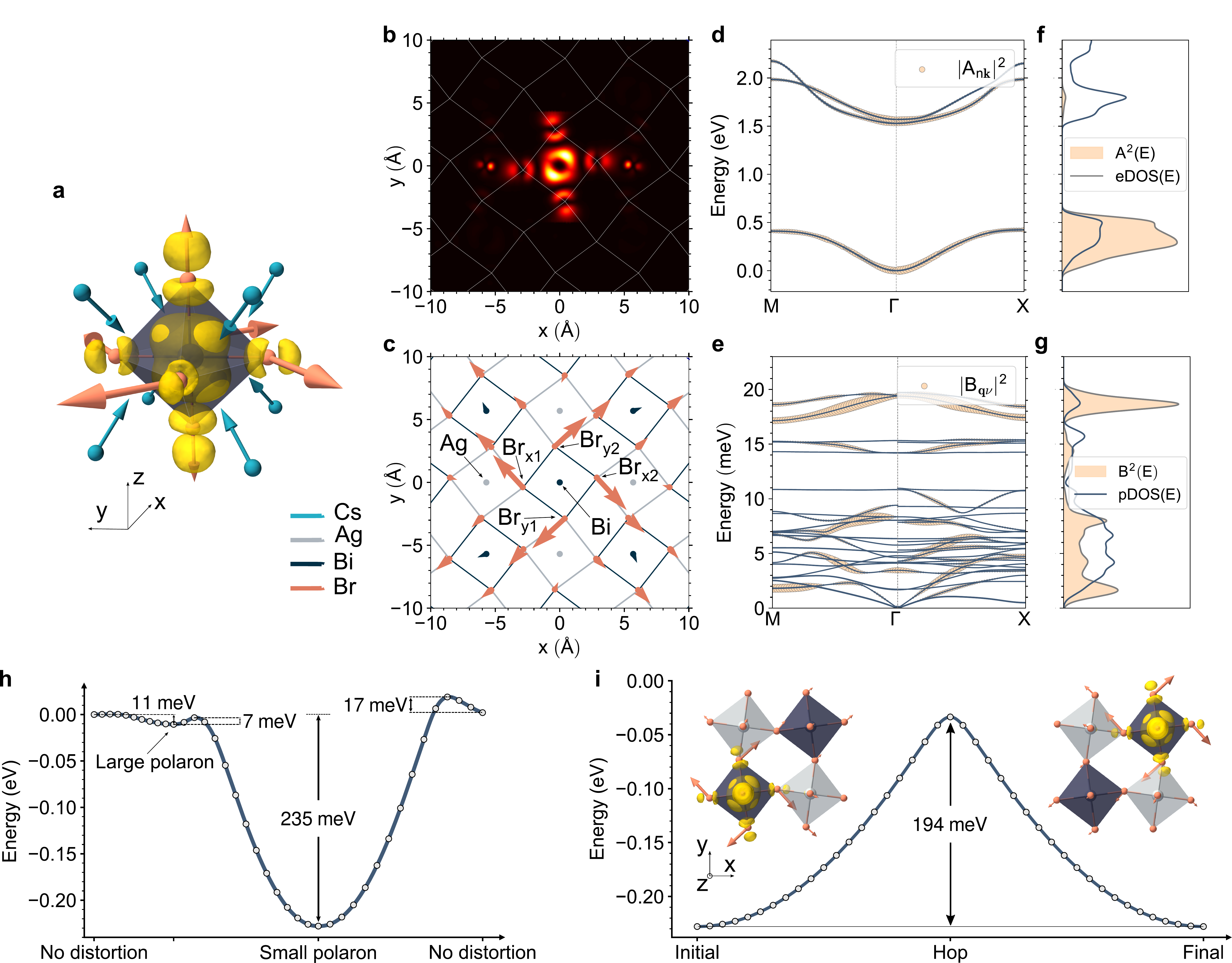}
    \caption{\textbf{Small ferrotoroidic electron polarons in halide perovskites.} \textbf{a} 3D view of the wavefunction of the small electron polaron in Cs$_2$AgBiBr$_6$ (surface) and its accompanying atomic displacements (arrows). \textbf{b} Cross-section of the square modulus of the polaron wave function on an $xy$ plane through the polaron center. \textbf{c} Projection of the atomic displacements of the polaron in (a) on an $xy$ plane through the polaron center. The color code is the same as in (a). \textbf{d} Analysis of the electron polaron in terms of Bloch wavefunctions. The square moduli of the spectral weigths $|A_{n\bf k}|^2$ shown in Eq.~(S4) of Supplementary Note~2 are superimposed on the electron band structure; the radius of the discs is proportional to the magnitude of each coefficient. The zero of the energy axis is referred to the conduction band bottom. \textbf{e} Same as (d), but for the phonon band structure and the coefficients $|B_{{\bf q}\nu}|^2$ shown in Eq.~(S5) of Supplementary Note~2. \textbf{f} Density of states of the spectral weights (filled orange area), together with the electron density of states (blue line).
    \textbf{g} Same as in (f), but for the phonon density of states. \textbf{h} Polaron energy landscape along a generalized coordinate that starts from the original undistorted configuration, goes through the large electron polaron configuration of Fig.~\ref{fig1}\textbf{g}, then through the small electron polaron configuration of Fig.~\ref{fig3}\textbf{a}, and back to the undistorted configuration. The energies of the potential wells and the transition barriers are indicated. \textbf{i} Polaron hopping barrier between nearest-neighbor Bi sites. Top views of the polaron wave function and its associated atomic displacements are shown as insets for the initial and final configurations.}
    \label{fig2}
\end{figure*}

Besides large electron and hole polarons, our search indicates that Cs$_2$AgBiBr$_6$ also hosts small electron polarons. Figure~\ref{fig2}\textbf{a} shows that the small polaron is localized around the Bi cations. This polaron induces an octahedral rotation whose displacement patterns retain the same helical Bloch point topology as the large polarons in Fig.~\ref{fig1}. In fact, also in this case we calculate a topological charge $Q=-1$, vorticity $q=1$ and helicity $\gamma=-\pi/4$, in line with what we found for the large polarons. This suggests that topologically nontrivial polarons are a general feature of halide perovskites. 

\begin{figure*}[ht!]
    \centering
    \includegraphics[width=0.8\textwidth]{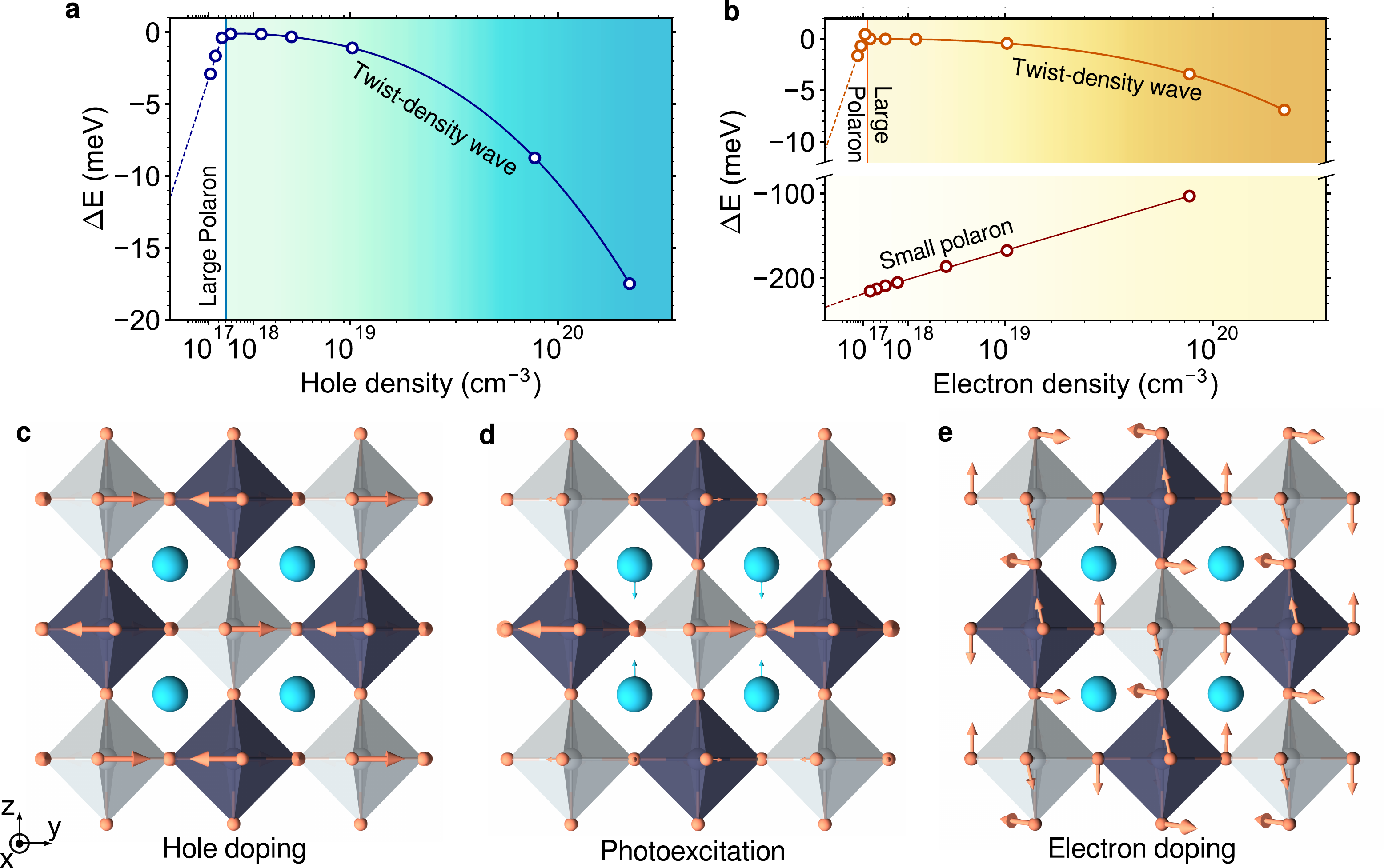}
    \caption{\textbf{Periodic twist-density waves in halide perovskites.} \textbf{a} Formation energy of hole polaron in Cs$_2$AgBiBr$_6$ as a function of hole density. The zero of the energy axis is set to the energy of the undistorted structure. The hole density is tuned by varying the Brillouin zone grid in Eqs.~(S6) and (S7) of Supplementary Note~2. Calculated values are shown as discs, the solid line is a guide to the eye. The vertical line marks the critical density for the Mott metal-insulator transition. The dashed line represents the extrapolation of the polaron energy to the limit of infinite supercell. \textbf{b} Same as (a), but for the electron polaron. In this case, the large polaron of Fig.~\ref{fig1} and the twist-density waves shown in (c) and (e) coexist with the energetically more stable small polaron of Fig.~\ref{fig2}. \textbf{c}, \textbf{e} Atomic displacements associated with twist density waves above the Mott density for holes and electrons, respectively. Hole doping leads to octahedral rotations around the $z$ axis, electron doping leads to octahedral tilts around the $[\bar{1}, 1, 0]$ direction. \textbf{d} Atomic displacements associated with a neutral photo-excitation, also showing a twist-density wave pattern (cf.\ Supplementary Note~5).}
    \label{fig3}
\end{figure*}

In Fig.~\ref{fig2}\textbf{c} we show that, in the presence of the small electron polaron, the equatorial Br anions define a circular motion around the central Bi cation. Since these ions carry non-vanishing Born effective charges, their displacements  generate microscopic electric dipoles (cf.\ Supplementary Fig.~S7). While the total electric dipole vanishes by symmetry, the polaron carries a finite toroidal moment around the polaron center \cite{Spaldin2008,Shimada2020}. This type of ferroic order has been investigated in the context of ferroelectrics as a potential building block for non-volatile ferroelectric random access memories \cite{Naumov2004}. Our computed ferrotoroidic moment for the small electron polaron in Cs$_2$AgBiBr$_6$ is $4\times 10^{-6}$~e/\AA, and points along the apical direction (Fig.~S7). This is a fairly low value when compared to the moments required for ferroelectric memories (e.g. 0.2~e/\AA, Ref.~\cite{Damodaran2017}). However, since the toroidal moments of different polarons all point in the same direction (Fig.~S8), and many small polarons can be packed together in this compound (Fig.~\ref{fig3}\textbf{b}), it should be possible to achieve technologically relevant ferrotoroidic moments at large polaron concentrations.

Figure~\ref{fig2}\textbf{b} shows the electron density and the atomic displacements of the small electron polaron within the equatorial plane. The electron density is strongly localized around a single Bi cation and the surrounding Br anions. This negative charge distribution causes the anions defining the BiBr$_6$ octahedron move away from the Bi cation, while the nearest-neighbor Cs cations move toward it. Figure~\ref{fig2}\textbf{d} shows that the polaron draws its weight primarily from the lowest conduction band with Bi-$6p$ and Br-$4p$ character; however, unlike for the large electron polaron (Fig.~S6), in this case the polaron is a superposition of states from the entire Brillouin zone. Figures~\ref{fig2}\textbf{e} and \ref{fig2}\textbf{g} show how the atomic displacements of the polaron are dominated by the octahedral breathing mode (18.6~meV) and the octahedral rotation mode (1.8~meV), while the modes around 8~meV contribute to the Cs displacements, in line with the arrows in Fig.~\ref{fig2}\textbf{a}.

\medskip

\subsection*{Polaron energetics and dynamics}

In Fig.~\ref{fig2}\textbf{h} we analyze the polaron energy landscape along a generalized coordinate that connects the undistorted perovskite structure to the configuration of the large polaron, and then to the structure of the small electron polaron (cf.~Supplementary Note~4). We set the zero of the energy axis to the perovskite structure without polarons; negative values in this plot indicate stable polarons. This figure shows how the small electron polaron is the most stable species, with a formation energy of 235~meV, and thus the most highly populated state at thermal equilibrium. However, note that an energy barrier must be overcome for an electron to localize into the small polaron starting from the fully-delocalized Bloch state (17~meV) or from the large polaron (7~meV). Conversely, the formation of the large polaron starting from the delocalized Bloch state is barrierless. As a result, upon photoexcitation, at short timescales and before reaching thermal equilibrium, electrons will initially form large polarons, and gradually convert into small polarons. During this transient, a mixed population of small and large polarons is possible. These results are in excellent agreement with the recent observation of two-time electron localization in Cs$_2$AgBiBr$_6$ upon photoexcitation \cite{Wu2021,Wright2021}, which was rationalized as the formation of a large polaron intermediate on ultrafast timescales, followed by a slow transition to a small polaron. This interpretation is further corroborated by the timescales of the phonons involved in this process (cf. Supplementary Fig.~S6): our calculated oscillation period of the longitudinal-optical phonon that drives the large polaron (0.21~ps) coincides with the initial fast localization time reported in Ref.~\citenum{Wu2021} (0.25~ps). Similarly, the period of the octahedral rotation that contributes to the small polaron (2.3~ps) falls between the slow localization times 1.0~ps and 4.7~ps reported in Refs.~\citenum{Wright2021} and \citenum{Wu2021}, respectively. Additional evidence for this scenario is provided by the experimental observation of displacive excitations of coherent phonons at $\sim$22~meV and $\sim$1~meV \cite{Wu2021}, which match our calculated energies of the octahedral rotation and breathing modes in Fig.~\ref{fig2}\textbf{e} and \ref{fig2}\textbf{g}.

Figure~\ref{fig2}\textbf{i} shows the energy landscape associated with the nearest-neighbor hopping of the small electron polaron. The high transition barrier (194~meV) suggests that this species is not mobile. In fact, an estimate of the polaron mobility via the Emin-Holstein-Austin-Mott theory \cite{Deskins2007,Emin2012} yields a very low value, $<$0.01~cm$^2$/Vs (cf.\ Supplementary Note~4). This finding is compatible with experimentally reported mobilities in the range of 1~cm$^2$/Vs \cite{Bartesaghi2018,Longo2020} or lower \cite{Wu2021}.

\medskip

\subsection*{Twist-density waves}

The polaronic species identified thus far correspond to the ideal limit of isolated polarons. To enable further comparison with experiments \cite{Kirschner2019,Zhang2023,Seiler2023}, in Fig.~\ref{fig3}\textbf{a} and \ref{fig3}\textbf{b} we analyze the phase diagram of these species as a function of carrier concentration. Upon increasing carrier concentration, nearby polarons begin to overlap, leading to complete wavefunction delocalization at the critical density for the Mott metal-insulator transition. From Fig.~\ref{fig3}\textbf{a} and \ref{fig3}\textbf{b} we obtain critical densities of $3\times 10^{17}$~cm$^{-3}$ and $10^{17}$~cm$^{-3}$ for hole and electron polarons, respectively. Both estimates are in good agreement with the recently reported critical density of $\sim\!\!10^{18}$~cm$^{-3}$ for lead halide perovskites at room temperature \cite{Zhang2023}. 
In Fig.~\ref{fig3}\textbf{b} we also see that the small electron polaron is stable over a very broad range of carrier densities. This finding agrees with the experimental observation that the localization dynamics in Cs$_2$AgBiBr$_6$ is not sensitive to photon fluence \cite{Wu2021}. 

Figures~\ref{fig3}\textbf{c} and \ref{fig3}\textbf{e} show yet another surprise in the metallic regime above the Mott critical density. In this regime, electron wavefunctions are fully delocalized, therefore one would expect the crystal to settle in its ground-state periodic structure. Instead, lattice-periodic distortions emerge, corresponding to rigid rotations of the octahedra. We name these configurations twist-density waves for their similarity with charge-density waves in low-dimensional materials \cite{Monceau2012}. 

These configurations can be rationalized in terms of the Goldschmidt tolerance factor \cite{Goldschmidt1926,Filip2018}. For simple perovskites with formula ABX$_3$, the tolerance factor is given by $t = (r_{\rm A}+r_{\rm X})/\sqrt{2}(r_{\rm B}+r_{\rm X})$, with $r_{\rm A}$, $r_{\rm B}$, $r_{\rm X}$ being the ionic radii. As the tolerance factor increases from $t=0.7$ to $t=1$, perovskites evolve from the orthorhombic to the cubic structure in order to maximize the packing density~\cite{Filip2018}. In Fig.~\ref{fig3}\textbf{c}, we see that electron removal induces octahedral rotations around the apical axis, which tend to restore the cubic structure; this is in agreement with the fact that electron removal effectively reduces the ionic radius of Ag, thereby increasing the tolerance factor. Conversely, in Fig.~\ref{fig3}\textbf{e}, we see that electron addition makes the apical axes tilt toward the orthorhombic structure; this is consistent with the increase of the ionic radius of Bi upon electron addition, which causes a reduction of the tolerance factor.

The structural sensitivity to the carrier density suggests that halide perovskites could be employed to investigate light-driven phase transitions. To test this hypothesis, in Fig.~\ref{fig3}\textbf{d} we consider Cs$_2$AgBiBr$_6$ in the presence of both electrons and holes, and we find that indeed photoexcitation drives octahedral rotations toward a cubic structure. This observation agrees with recent experiments showing that high-fluence photo-doping of methylammonium lead iodide induces octahedral rotations \cite{Wu2017}, leading to a non-equilibrium pseudo-cubic phase \cite{Panuganti2023,Leonard2023}. Similar photo-induced octahedral rotations have been reported in two-dimensional halide perovskites \cite{ZhangH2023}.

\medskip

\section*{Discussion}

The diversity of polaronic species in halide perovskites identified here provides a unified conceptual framework to rationalize the multitude of experimental observations of large polarons \cite{Miyata2017_2,Guzelturk2021}, small polarons \cite{Luo2018,Wu2021}, and photo-induced structural rearrangements \cite{Leonard2023,ZhangH2023} in these materials. Furthermore, the unexpected topological nature of these quasiparticles might help explain the unique photophysical properties of these compounds in terms of topological protection against scattering and recombination~\cite{Qian2023} (cf. Supplementary Note 6), and could lead to novel and unique signatures
in ultrafast electron diffraction experiments (cf.
Supplementary Note 7 and Fig. S10).
We expect that thermal and zero-point vibrations will lead to a broadening of optical lineshapes and diffraction patterns \cite{Lafuente2022}, which are distinct and complementary effects to those induced by the polaron localization and the associated topological charge discussed herein.

The present study prompts the question on whether related electronic phases could emerge in other families of perovskites. Photo-induced octahedral rotations have been reported in SrTiO$_3$ \cite{Porer2018} and EuTiO$_3$ \cite{Porer2019}; polar vortices have been observed in PbTiO$_3$/SrTiO$_3$ superlattices \cite{Yadav2016,Damodaran2017}; ferroelectric skyrmions, merons and Bloch-point structures have been reported in oxide superstructures \cite{Das2019,Wang2020,Han2022,Junquera2023}; and ferrotoroidic polarons have been theoretically discussed for SrTiO$_3$ \cite{Shimada2020}. 
Beyond perovskites, the concept of topological polarons has also been proposed in the context of cold-atom experiments \cite{Grusdt2016, Grusdt2019}, suggesting that these quasiparticles may be more common than previously thought.
While these diverse observations have been made in different contexts and appear unrelated to one another, our findings lead us to speculate that they could all be system-specific realizations of one or more new classes of polarons with topological properties.

\medskip

\acknowledgments

The authors are grateful to Joshua Leveillee for insightful discussions during the initial stages of this work.
This research is supported by the Computational Materials Sciences Program funded by the U.S. Department of Energy, Office of Science, Basic Energy Sciences, under Award No. DE-SC0020129. This research used resources of the National Energy Research Scientific Computing Center, a DOE Office of Science User Facility supported by the Office of Science of the U.S.  Department of Energy under Contract No. DE-AC02-05CH11231. The authors also acknowledge the Texas Advanced Computing Center (TACC) at The University of Texas at Austin for providing additional HPC resources, including the Frontera and Lonestar6 systems, that have contributed to the research results reported with in this paper. URL: http://www.tacc.utexas.edu.


\clearpage
\newpage

\renewcommand\theequation{S\arabic{equation}}
\renewcommand{\thesection}{\arabic{section}}
\renewcommand{\thetable}{S\arabic{table}}
\renewcommand\thefigure{S\arabic{figure}}
\titleformat{\section}{\large\bfseries}{\thesection.}{5pt}{}
\renewcommand{\baselinestretch}{1.15}
\setlength{\parindent}{0pt}
\setlength{\parskip}{3pt}
\titlespacing{\section}{0pt}{16pt}{5pt}
\def\bibsection{\section*{References}}
\setcounter{equation}{0}
\setcounter{figure}{0}

\onecolumngrid

\onecolumngrid

\begin{center}
\textbf{\large Supplementary Information for:\\[4pt] Topological polarons in halide perovskites}
\end{center}

\vspace*{5pt}
Contents:\\

Supplementary Notes 1-7\\
Supplementary Figures S1-S11 

\section*{Supplementary Note 1: Computational details} \label{sec:supp1}

All \textit{ab initio} density-functional theory (DFT) calculations are performed using the Quantum ESPRESSO package \cite{Giannozzi2017} (electronic structure and lattice vibrational properties), the Wannier90 code \cite{Pizzi2020} (maximally-localized Wannier functions) and the EPW code \cite{Ponce2016,Lee2023} (interpolation of electron-phonon matrix elements and polarons).  We describe Cs$_2$AgBiBr$_6$ within the Perdew-Burke-Ernzerhof generalized gradient approximation \cite{Perdew1996}, using optimized norm-conserving Vanderbilt (ONCV) pseudopotentials \cite{Hamann2013,vanSetten2018} and plane waves with a kinetic energy cutoff of 100 Ry.  We consider the low-temperature tetragonal structure of Cs$_2$AgBiBr$_6$ with space group $I4/m$.  Our optimized lattice parameters are $a=8.04$~\AA\ and $c=11.63$~\AA\ \cite{Leveillee2021}, in good agreement with X-ray diffraction measurements at 30~K \cite{Schade2019}.  Phonon frequencies and electron-phonon matrix elements are computed within density functional perturbation theory \cite{Baroni2001}.  Coarse momentum grids of 6$\times$6$\times$6 $\mathbf{k}$ and $\mathbf{q}$-points are used for the ground state electron and lattice dynamics calculations, respectively.  Spin-orbit coupling is important for the description of conduction band, and is included in the calculations for electron polarons.  Electron energies, phonon frequencies, and electron-phonon matrix elements are interpolated to dense grids by means of Wannier-Fourier interpolation \cite{Marzari1997,Giustino2007}.  In the process of obtaining maximally localized Wannier functions, the subset formed by the 6 lowest conduction bands and 26 highest valence bands are considered for electron and hole polarons, respectively.  The method presented in Ref.~\citenum{Verdi2015} is used to deal with the long-range part of the electron-phonon matrix element for polar materials.

To correctly describe an isolated polaron in the crystal, we solve the \textit{ab initio} polaron equations described in Supplementary Note~2 with increasingly denser $\mathbf{k}$ and $\mathbf{q}$ meshes, and extrapolate the formation energy to the limit of infinite supercell \cite{Sio2019_b}.  Inhomogeneous reciprocal-space meshes are used to describe polarons in real-space non-diagonal supercells, as described in Supplementary Note~3. In particular, momentum grids corresponding to $N \times N \times N$ parent cubic supercells, with $N = \{ 2, 4, 6, 8, 12, 14, 16, 18 \}$ for hole polarons and $N = \{ 2, 4, 8, 12, 16, 18, 20, 22 \}$ for electron polarons are used in Fig.~3 of the main text. The largest supercell considered contains $22 \times 22 \times 22 \times 40 = 425,920$ atoms. Fig.~\textbf{1h} of the main text shows 32,000 Ag atoms within a $20 \times 20 \times 20$ pseudocubic supercell.

\section*{Supplementary Note 2: \textit{Ab initio} polaron equations} \label{sec:supp2}

Within the \textit{ab initio} theory of polarons developed in Refs.~\citenum{Sio2019,Sio2019_b}, the formation energy of the polaron is written as a self-interaction-free functional of the polaron wave function $\psi(\mathbf{r})$ and the associated atomic displacements $\Delta \boldsymbol{\tau}$:
\begin{equation} \label{eq:ef}
  \Delta E_{\mathrm{f}}
  = 
  \int d\mathbf{r} \psi^{*}(\mathbf{r}) \hat{H}_{\mathrm{KS}}^{0} \psi(\mathbf{r})
  +
  \sum_{\kappa\alpha p} \int d\mathbf{r} \frac{\partial V_{\mathrm{KS}}^{0}}{\partial \tau_{\kappa\alpha p}}
  |\psi(\mathbf{r})|^2 \Delta \tau_{\kappa\alpha p}
  +
  \frac{1}{2} \sum_{\kappa\alpha p \\ \kappa'\alpha' p'} 
  C^{0}_{\kappa\alpha p, \kappa'\alpha' p'} \Delta \tau_{\kappa\alpha p} \Delta \tau_{\kappa'\alpha' p'} ~,
\end{equation}
where $\hat{H}_{\mathrm{KS}}^{0}$, $V_{\mathrm{KS}}^{0}$ and $C^{0}_{\kappa\alpha p, \kappa'\alpha' p'}$ are the Kohn-Sham Hamiltonian, Kohn-Sham potential, and interatomic force constant matrix in the ground state configuration without the polaron, respectively. $\Delta \boldsymbol{\tau}_{\kappa p}$ indicates a displacement of the atom $\kappa$ in the unit cell $p$ along the cartesian direction $\alpha$, and the integrals are taken over the Born-von Karman (BvK) supercell formed by $N_p$ unit cells. Upon variational minimization with respect to the wave function and displacements, the following set of coupled equations are obtained:
\begin{equation} \label{eq:var1}
  \hat{H}_{\mathrm{KS}}^{0} \, \psi(\mathbf{r})
  +
  \sum_{\kappa\alpha p} \frac{\partial V_{\mathrm{KS}}^{0}}{\partial \tau_{\kappa \alpha p}}
  \Delta \tau_{\kappa \alpha p} \, \psi(\mathbf{r})
  =
  \varepsilon \, \psi(\mathbf{r}) ~,
\end{equation}
\begin{equation} \label{eq:var2}
  \Delta \tau_{\kappa \alpha p}
  =
  -\sum_{\kappa'\alpha' p'} (C^{0})^{-1}_{\kappa\alpha p,\kappa'\alpha'p'}
  \int d\mathbf{r} \, \frac{\partial V_{\mathrm{KS}}^{0}}{\partial \tau_{\kappa'\alpha'p'}}
  \, |\psi(\mathbf{r})|^2 ~,
\end{equation}
where $\varepsilon$ is the Lagrange multiplier for the normalization constraint on the wave function, and can be interpreted as the quasiparticle excitation energy of the polaron \cite{Lafuente2022,Lafuente2022_b}.

By expressing the polaron wave function as a linear combination of Kohn-Sham states $\psi_{n\mathbf{k}}$ with eigenvalues $\varepsilon_{n\mathbf{k}}$:
\begin{equation} \label{eq:exp1}
  \psi(\mathbf{r}) = \frac{1}{\sqrt{N_p}}
  \sum_{n\mathbf{k}} A_{n\mathbf{k}} \psi_{n\mathbf{k}} ~,
\end{equation}
and writing the atomic displacements as linear combinations of phonon eigenmodes $e_{\kappa\alpha,\nu}(\mathbf{q})$ with frequencies $\omega_{\mathbf{q}\nu}$:
\begin{equation} \label{eq:exp2}
  \Delta\tau_{\kappa\alpha p } = -\frac{2}{N_p}
  \sum_{\mathbf{q}\nu} B^{*}_{\mathbf{q}\nu} 
  \left( \frac{\hbar}{2M_\kappa \omega_{\mathbf{q}\nu}} \right)^{1/2}
  e_{\kappa\alpha,\nu}(\mathbf{q}) e^{i\mathbf{q}\cdot\mathbf{R}_p},
\end{equation}
where $M_\kappa$ is the mass of the atom $\kappa$ and $\mathbf{R}_p$ is the lattice vector of the unit cell $p$, \eqref{eq:var1} and \eqref{eq:var2} are rewritten as a coupled system of equations for the coefficients $A_{n\mathbf{k}}$ and $B_{\mathbf{q}\nu}$:
\begin{equation} \label{eq:vark1}
  \frac{2}{N_p} \sum_{\mathbf{q}m\nu} B_{\mathbf{q}\nu}
  \, g_{mn\nu}^{*}(\mathbf{k},\mathbf{q}) \, A_{m\mathbf{k+q}}
  =
  (\varepsilon_{n\mathbf{k}}-\varepsilon) A_{n\mathbf{k}} ~,
\end{equation}
\begin{equation} \label{eq:vark2}
  B_{\mathbf{q}\nu} = \frac{1}{N_p}
  \sum_{mn\mathbf{k}} A^{*}_{m\mathbf{k+q}}
  \frac{g_{mn\nu}(\mathbf{k},\mathbf{q})}{\hbar\omega_{\mathbf{q}\nu}} A_{n\mathbf{k}} ~,
\end{equation}
which are referred to as the \textit{ab initio} polaron equations. These equations constitute the DFT approximation of the more general many-body polaron equations introduced in Refs.~\citenum{Lafuente2022,Lafuente2022_b}.

For electron polarons, the formation energy in the polaronic ground state is obtained by combining \eqref{eq:ef}, \eqref{eq:exp1}, \eqref{eq:exp2}, \eqref{eq:vark1} and \eqref{eq:vark2}:
\begin{equation} \label{eq:efmin}
  \Delta E_{f}
  =
  \frac{1}{N_p} \sum_{n\mathbf{k}} |A_{n\mathbf{k}}|^2
  (\varepsilon_{n\mathbf{k}}-\varepsilon_{\mathrm{CBM}})
  -
  \frac{1}{N_p} \sum_{\mathbf{q}\nu} |B_{\mathbf{q}\nu}|^2 \hbar\omega_{\mathbf{q}\nu} ~,
\end{equation}
where $\varepsilon_{\mathrm{CBM}}$ is the Kohn-Sham eigenvalue at the conduction band minimum.  For hole polarons, $\varepsilon_{\mathrm{CBM}}$ is replaced by the eigenvalue of the valence band maximum, and the first term in the righ-hand side of \eqref{eq:efmin} acquires a minus sign.

For visualization purposes, a more convenient real-space representation of the polaron wave function is obtained by using a linear combination of maximally localized Wannier functions $w_m({\bf R}_p)$ \cite{Marzari1997}, which are related to Kohn-Sham states by:
\begin{equation} \label{eq:b2w}
  \psi_{n\mathbf{k}} = 
  \frac{1}{\sqrt{N_p}} 
  \sum_{mp} e^{i\mathbf{k}\cdot\mathbf{R}_p}
  U^{\dagger}_{mn\mathbf{k}}
  w_m(\mathbf{r}-\mathbf{R}_p) ~.
\end{equation}
Here, ${\bf R}_p$ denotes a vector of the direct lattice, and $U^{\dagger}_{mn\mathbf{k}}$ is the unitary matrix that generates the smooth Bloch gauge.  Replacing \eqref{eq:b2w} in \eqref{eq:exp1}, and defining
\begin{equation}
  A_{m}(\mathbf{R}_p)
  =
  \frac{1}{N_p} 
  \sum_{n\mathbf{k}} e^{i\mathbf{k}\cdot\mathbf{R}_p}
  U^{\dagger}_{mn\mathbf{k}}
  A_{n\mathbf{k}} ~,
\end{equation}
we obtain:
\begin{equation}
  \psi(\mathbf{r}) = 
  \sum_{mp} A_{m}(\mathbf{R}_p) w_m(\mathbf{r}-\mathbf{R}_p) ~.
\end{equation}
This representation allows us to interpret $|A_{m}(\mathbf{R}_p)|^2$ as the spatial envelope of the polaron wave function. This envelope is useful to visualize the large polarons in Figs.~\textbf{1b} and \textbf{1c} without rendering the atomic-scale details contained in the Wannier functions.  

\section*{Supplementary Note 3: Polarons in non-diagonal supercells} \label{sec:supp3}

In this note we describe the methodology for computing polarons in orthogonal BvK supercells when the primitive lattice vectors are not orthogonal. In this case of Cs$_2$AgBiBr$_6$ considered here, this procedure is useful to analyze polaron wavefunctions and atomic displacements using a pseudocubic BvK supercell.


We denote by $\mathbf{a}_{s1}$, $\mathbf{a}_{s2}$, $\mathbf{a}_{s3}$ the primitive lattice vectors of a BvK supercell, and by $\mathbf{a}_{p1}$, $\mathbf{a}_{p2}$, $\mathbf{a}_{3}$ the primitive lattice vectors of the crystal unit cell. These vectors must be related as follows:
\begin{equation}
  \begin{pmatrix}
  \mathbf{a}_{s1}\\
  \mathbf{a}_{s2}\\
  \mathbf{a}_{s3}
  \end{pmatrix}
  =
  \begin{pmatrix}
  S_{11} & S_{12} & S_{13}\\
  S_{21} & S_{22} & S_{23}\\
  S_{31} & S_{32} & S_{33}
  \end{pmatrix}
  \begin{pmatrix}
  \mathbf{a}_{p1}\\
  \mathbf{a}_{p2}\\
  \mathbf{a}_{p3}
  \end{pmatrix} ~, \label{eq.SC1}
\end{equation}
where the matrix elements $S_{ij}$ are integers. The transformation between the corresponding reciprocal lattice vectors is:
\begin{equation}
  \begin{pmatrix}
  \mathbf{b}_{s1}\\
  \mathbf{b}_{s2}\\
  \mathbf{b}_{s3}
  \end{pmatrix}
  =
  \begin{pmatrix}
  \bar{S}_{11} & \bar{S}_{12} & \bar{S}_{13}\\
  \bar{S}_{21} & \bar{S}_{22} & \bar{S}_{23}\\
  \bar{S}_{31} & \bar{S}_{32} & \bar{S}_{33}
  \end{pmatrix}
  \begin{pmatrix}
  \mathbf{b}_{p1}\\
  \mathbf{b}_{p2}\\
  \mathbf{b}_{p3}
  \end{pmatrix}\label{eq.SC2}
\end{equation}
where $\mathbf{b}_{s1}$, $\mathbf{b}_{s2}$, $\mathbf{b}_{s3}$ are the primitive reciprocal lattice vectors of the BvK supercell, and $\mathbf{b}_{p1}$, $\mathbf{b}_{p2}$, $\mathbf{b}_{3}$ are the primitive vectors of the reciprocal lattice of the crystal unit cell. The matrix elements of the two transformations defined in \eqref{eq.SC1} and \eqref{eq.SC2} are related by $\bar{S}=(S^{-1})^{T}$~\cite{Lloyd2015}.

When the off-diagonal elements of the matrix $S$ in \eqref{eq.SC1} are non-zero, one obtains a so-called ``non-diagonal'' BvK supercell~\cite{Lloyd2015}. A non-diagonal supercell differs from a standard supercell in that it does not have the same shape as the primitive unit cell. This observation was exploited in Ref.~\citenum{Lloyd2015} to generate supercells for computing phonons at desired $\bf q$-vectors from finite-difference calculations.
Here, we employ the same strategy as in Ref.~\citenum{Lloyd2015} but in reverse: first we choose the shape of the BvK supercell, then we use \eqref{eq.SC2} to determine the Brillouin zone grid that we need in \eqref{eq:vark1}-\eqref{eq:vark2} to obtain polarons in such a supercell.

The primitive lattice vectors of the body-centered tetragonal unit cell of Cs$_2$AgBiBr$_6$ are~\cite{Mehl2017}:
\begin{equation}
  \begin{cases}
    \mathbf{a}_{p1} = (-a\hat{\mathbf{x}} + a\hat{\mathbf{y}} + c\hat{\mathbf{z}})/2, \\
    \mathbf{a}_{p2} =  (\phantom{-}a\hat{\mathbf{x}} - a\hat{\mathbf{y}} + c\hat{\mathbf{z}})/2 ,\\
    \mathbf{a}_{p3} =  (\phantom{-}a\hat{\mathbf{x}} + a\hat{\mathbf{y}} - c\hat{\mathbf{z}})/2 ,
  \end{cases}
\end{equation}
where $\hat{\mathbf{x}}, \hat{\mathbf{y}}, \hat{\mathbf{z}}$ are the unit vectors along the Cartesian axes, and $a$ and $c$ are the lattice parameters. To compute polarons, we choose a pseudocubic supercell with primitive lattice vectors:
\begin{equation}
  \begin{cases}
    \mathbf{a}_{s1} = N(a\hat{\mathbf{x}} + a\hat{\mathbf{y}}), \\
    \mathbf{a}_{s2} = N(a\hat{\mathbf{x}} - a\hat{\mathbf{y}}), \\
    \mathbf{a}_{s3} = Nc\hat{\mathbf{z}},
  \end{cases}
\end{equation}
where $N$ is a positive integer. Such a supercell is obtained by using \eqref{eq.SC1} with the transformation matrix:
\begin{equation}
  S
  =
  N \begin{pmatrix}
     \phantom{-}1 & 1 & 2 \\
    -1 & 1 & 0 \\
     \phantom{-}1 & 1 & 0
  \end{pmatrix} ~,
\end{equation}
which contains $|\mathrm{det}\, S| = 4N$ crystal unit cells. The inverse transformation $\bar S$ is given by:
\begin{equation}
  \bar S
  =
  \frac{1}{2N} \begin{pmatrix}
     \phantom{-}0 & \phantom{-}0 & \phantom{-}1 \\
     -1 & \phantom{-}1 & \phantom{-}0 \\
     \phantom{-}1 & \phantom{-}1 & -1
  \end{pmatrix} ~.
\end{equation}
The $\bf k$-point (and $\bf q$-point) grids to be used in the solution of \eqref{eq:vark1} and \eqref{eq:vark2} are given by the points
${\bf k} = i\mathbf{b}_{s1}+j\mathbf{b}_{s2}+k\mathbf{b}_{s3}$
that fall within the first Brillouin zone of the crystal unit cell,
being $\mathbf{b}_{s1}$, $\mathbf{b}_{s2}$, and $\mathbf{b}_{s3}$ the BvK supercell reciprocal lattice vectors obtained from \eqref{eq.SC2}.

For ease of visualization, in all figures of the main text the Cartesian axes are aligned with the orthogonal axes of the pseudocubic unit cell.

\section*{Supplementary Note 4: Polaron energy landscape} \label{sec:supp4}

The dependence of the energy functional in \eqref{eq:ef} on the atomic displacements allows us to explore the adiabatic potential energy surface of the polaron. The ground state energy of the polaron for a given configuration of the atomic displacements, $\{\Delta \tau_{\kappa\alpha p}\}$, is obtained  by performing the variational minimization of the energy functional with respect to the polaron wave function, leading to \eqref{eq:var1}. Substituting \eqref{eq:exp1} and \eqref{eq:exp2} in \eqref{eq:var1} yields the following eigenvalue equation \cite{Sio2019_b}:
\begin{equation} \label{eq:eig}
  \sum_{n'\mathbf{k}'}
  H_{n\mathbf{k}, n'\mathbf{k}'} A_{n'\mathbf{k}'}
  =
  \varepsilon A_{n\mathbf{k}} ~,
\end{equation}
where 
\begin{equation} \label{eq:ham}
  H_{n\mathbf{k}, n'\mathbf{k}'}
  =
  \delta_{n\mathbf{k}, n'\mathbf{k}'} \varepsilon_{n\mathbf{k}}
  - \frac{2}{N_p} 
  \sum_{\nu} B^{*}_{\mathbf{k}-\mathbf{k}' \nu} 
  \, g_{nn'\nu}(\mathbf{k}', \mathbf{k}-\mathbf{k}').
\end{equation}
An explicit relation between the eigenvalue in \eqref{eq:eig} and the formation energy can be obtained by using the definition of the interatomic force constant matrix in terms of the phonon eigenmodes \cite{Giustino2017}, and substituting \eqref{eq:var1} and \eqref{eq:exp2} in \eqref{eq:ef}:
\begin{equation} \label{eq:ef_eigval}
  \Delta \tilde E_{\mathrm{f}} = \varepsilon - \varepsilon_{\mathrm{CBM}}
                          + \frac{1}{N_p} \sum_{\mathbf{q}\nu} 
                          |B_{\mathbf{q}\nu}|^2 \hbar \omega_{\mathbf{q}\nu} ~.
\end{equation}
In this expression, we use the tilde symbol in $\Delta \tilde E_{\mathrm{f}}$ to emphasize that this energy is a minimum with respect to the eletronic degrees of freedom, but the forces on the atoms are generally nonvanishing in this configuration. If we minimize this quantity also with respect to the atomic displacements, then we obtain the formation energy $\Delta E_{\mathrm{f}}$ of \eqref{eq:efmin}.

In order to explore polaron energy surfaces we proceed as follows: First, we define a given displacement configuration $\{\Delta \tau_{\kappa\alpha p}\}$; then, we perform the transformation from $\{\Delta \tau_{\kappa\alpha p}\}$ to $\{ B_{\mathbf{q}\nu}\}$ using \eqref{eq:exp2}; we diagonalize the effective Hamiltonian in \eqref{eq:eig}; and we obtain the associated formation energy using \eqref{eq:ef_eigval}. 

Each section of the path along the configuration space used in Fig.~\textbf{2h} of the main text is defined by performing a linear interpolation between the initial and final displacement configurations. The horizontal axis is scaled by the distance between configurations along each section of the path. To define such a distance, we use the L1 norm of the difference vector, $d = \sum_{\kappa p} |\Delta \boldsymbol{\tau}^{\mathrm{final}}_{\kappa p} - \Delta \boldsymbol{\tau}^{\mathrm{initial}}_{\kappa p}|$.

In Fig.~\textbf{2i} of the main text, the final configuration is obtained by shifting the original small polaron configuration by one unit cell along the [1, 1, 0] direction, and a linear interpolation is used to define the hopping path.

In order to determine whether the polaron hopping is an adiabatic or a diabatic process, we consider the transmission coefficient $\kappa$ \cite{Deskins2007}. A value of $\kappa$ near 1 indicates adiabatic transfer, while a value near 0 indicates diabatic transfer. The transmission coefficient is given by:
\begin{equation} \label{eq:transcoef}
	\kappa = \frac{2 P }{1+P}~,
\end{equation}
where $P$ is the Landau-Zener transition probability for a potential energy surface crossing event, which is given by:
\begin{equation} \label{eq:lzprob}
	P = 1 - \mathrm{exp}\left[-\frac{V^2_{\mathrm{AB}} \, \pi^{3/2}}{h\nu_n \left(\lambda k_B T\right)^{1/2}}\right] ~.
\end{equation}
In this expression, $V_{\mathrm{AB}}$ is the electronic coupling between the initial and final states,
$\lambda$ is the reorganization energy,
$\nu_{n}$ is an effective frequency for nuclear motion, here taken to be the frequency of the longitudinal-optical phonon mode (4.7 THz), and $k_{\mathrm{B}}$ is the Boltzmann constant.
In order to extract $V_{\mathrm{AB}}$ and $\lambda$ from our calculations, 
we fit the minima of the potential energy surface with a parabola, as shown in Fig.~\ref{fig9}.
From this analysis, we get the values $V_{\mathrm{AB}}=24$ meV and $\lambda=874$ meV. 
Using these values in Eqs.~\ref{eq:transcoef} and \ref{eq:lzprob}, we obtain $\kappa=0.8$ at room temperature.
This prompts us to apply the adiabatic polaron transfer rate to estimate the polaron mobility as follows.

From the adiabatic hopping energy barrier $\Delta E_{\mathrm{hop}}$, the rate of polaron transfer at a given temperature $T$ can be obtained by \cite{Deskins2007}:
\begin{equation} \label{eq:transfer}
  k_{\mathrm{et}} = \nu_{n} \exp\left( \frac{-\Delta E_{\mathrm{hop}}}{k_{\mathrm{B}}T} \right) ~,
\end{equation}
From \eqref{eq:transfer}, the diffusion coefficient can be calculated using:
\begin{equation}
  D = R^2 \,n\, k_{\mathrm{et}} ~,
\end{equation}
where $R$ is the distance between transfer sites and $n$ is the number of neighboring electronic accepting sites \cite{Deskins2007}. The mobility is then estimated from the diffusion coefficient using Einstein's relation:
\begin{equation}
  \mu = \frac{e D}{k_{\mathrm{B}}T} ~,
\end{equation}
where $e$ is the electron charge. 

\section*{Supplementary Note 5: Supercell DFT calculations of the twist-density waves} \label{sec:supp5}

The twist-density wave configurations illustrated in Fig.~\textbf{3} of the main text are obtained as solutions of the polaron equations \eqref{eq:vark1}-\eqref{eq:vark2}. Since in this case the charge density of the extra electron or hole is lattice-periodic, it is also possible to perform direct DFT calculations of these configurations and cross-compare.

For this comparison, we employ $2 \times 2 \times 2$ diagonal supercells of the primitive unit cell of the crystal. The primitive cell is tetragonal and contains 20 atoms, therefore the supercell calculations contain 160 atoms and are manageable without any difficulty. By adding or removing one electron in this supercell, we achieve an electron or hole density of $1.66 \times 10^{20} ~\mathrm{e}/\mathrm{cm}^{3}$, which corresponds to the density of the rightmost data point in Fig.~\textbf{3a} and \textbf{3b}. 

For both cases of electron doping and hole doping, we perform three calculations: (i) ground state DFT calculation in the original undistorted structure; (ii) ground state DFT calculation in the structure obtained from the solution of \eqref{eq:vark1}-\eqref{eq:vark2}; (iii) DFT structural optimization starting from (ii). In all cases, we find that the total energy obtained in (ii) is lower than the total energy obtained in (i). Furthermore, the structural optimization performed in (iii) always leads to a stable solution where the atomic displacements remain close to the solution of \eqref{eq:vark1}-\eqref{eq:vark2}. This test provides an independent validation of the reliability of the twist-density wave configurations identified in this work.

In Fig.~\textbf{3d} of the main text, we consider a neutral photo-excitation using explicit DFT calculations. To this end, we perform $\Delta$-SCF calculations in the same supercell described above, by constraining the ground state to be a spin triplet. This choice leaves one hole in the valence bands and one electron in the conduction bands, and a vanishing net charge.

\section*{Supplementary Note 6: Reduction in the electron-hole recombination rate induced by the polaron localization} \label{sec:supp6}

In the following, we perform a quantitative estimate of the impact of polaron formation on the electron-hole recombination rate.

Given that the band gap in Cs$_2$AgBiBr$_6$ is indirect, the relevant recombination rate is the one connecting the conduction band minimum at $\mathbf{k}_{\mathrm{CBM}}$ and the valence band maximum $\mathbf{k}_{\mathrm{VBM}}$ via a phonon with momentum $\mathbf{q}=\mathbf{k}_{\mathrm{CBM}}-\mathbf{k}_{\mathrm{VBM}}$
Within second-order perturbation theory,
the transition rate mediated by a phonon $\nu$ with momentum $\mathbf{q}$ is given by 
\begin{equation}
	\begin{array}{lll}{W}_{fi}({{{\bf{q}\nu }}};\omega ) &\propto&\mathop{\sum}\limits_{\beta =\pm 1}{\left\vert {{{\bf{e}}}}\cdot [{{{{\bf{S}}}}}_{1,fi\nu }({{{\bf{q}}}})+{{{{\bf{S}}}}}_{2,fi\nu \beta }({{{\bf{q}}}})]\right\vert }^{2}\\ && \times \,\delta ({E }_{f}-{E }_{i}-\hbar \omega +\beta \hbar {\omega }_{{{{\bf{q}}}}\nu }) \,\end{array}
\end{equation}
where $\beta=\pm 1$ represent phonon emission and absorption processes, respectively, 
and the transition amplitudes are given by:
\begin{equation}
	{{{{\bf{S}}}}}_{1,fi\nu }({{{\bf{q}}}})=\mathop{\sum}\limits_{j}\frac{{g}_{fj\nu }({{{\bf{q}}}})\,{{{{\bf{v}}}}}_{ji}}{{E }_{j}-{E }_{i}-\hbar \omega +i\eta }~,
\end{equation}
and 
\begin{equation}
	{{{{\bf{S}}}}}_{2,fi\nu \beta }({{{\bf{q}}}})=\mathop{\sum}\limits_{j}\frac{{{{{\bf{v}}}}}_{fj}\,{g}_{ji\nu }({{{\bf{q}}}})}{{E }_{j}-{E }_{i}+\beta \hbar {\omega }_{{{{\bf{q}}}}\nu }+i\eta }~.
\end{equation}
In these expressions,
$g_{fi\nu}(\mathbf{q}) = \langle \Psi_f | \delta_{\mathbf{q}\nu} V | \Psi_i \rangle$ are the electron-phonon matrix elements and $\langle \Psi_f | \hat{\mathbf{v}} | \Psi_i \rangle$ are the matrix elements of the velocity operator.
Expanding the initial and final states in the Kohn-Sham basis:
\begin{equation}
	|\Psi_i\rangle = \frac{1}{\sqrt{N_p}} \sum_{n\mathbf{k}}^{\mathrm{val}} A_{n\mathbf{k}} |\psi_{n\mathbf{k}}\rangle ~,
\end{equation}
\begin{equation}
	|\Psi_f\rangle = \frac{1}{\sqrt{N_p}} \sum_{m\mathbf{k}'}^{\mathrm{cond}} A_{m\mathbf{k}'} |\psi_{m\mathbf{k}'}\rangle ~,
\end{equation}
and using $\langle \Psi_{\mathbf{k}'} | \delta_{\mathbf{q}\nu} V | \Psi_{\mathbf{k}} \rangle = \delta_{\mathbf{k}', \mathbf{k}+\mathbf{q}} \langle \Psi_{\mathbf{k}+\mathbf{q}} | \delta_{\mathbf{q}\nu} V | \Psi_{\mathbf{k}} \rangle$, we have:
\begin{equation}
	\begin{array}{lll}{W}_{fi}({{{\bf{q}\nu}}};\omega ) &\propto&\mathop{\sum}\limits_{\beta =\pm 1}{\left\vert \, \frac{1}{N_p} \,\sum_{n}^{\mathrm{val}}\sum_{m}^{\mathrm{cond}}\sum_{\mathbf{k}} A_{m\mathbf{k+q}}^{*} A_{n\mathbf{k}} \, ({{{\bf{e}}}}\cdot [{{{{\bf{S}}}}}_{1,mn\nu }({{{\bf{k}}}},{{{\bf{q}}}})+{{{{\bf{S}}}}}_{2,mn\nu \beta }({{{\bf{k}}}},{{{\bf{q}}}})]) \, \right\vert }^{2}\\ && \times \,\delta ({E }_{f}-{E }_{i}-\hbar \omega +\beta \hbar {\omega }_{{{{\bf{q}}}}\nu })~. \end{array}
\end{equation}
For simplicity, we will assume that only the lowest conduction band ($c$) has a significant weight in the electron polaron expansion $A_{m\mathbf{k+q}}=\delta_{m, c}A_{c\mathbf{k+q}}$, and similarly the highest valence band ($v$) in the hole polaron expansion $A_{n\mathbf{k}}=\delta_{n, v}A_{v\mathbf{k}}$, so that the summation over bands can be dropped. 
Moreover, we assume that the ${\bf{S}}_{cv\nu }({\bf{k}},{\bf{q}})$ matrices do not vary with $\mathbf{k}$, as only a narrow region around the band edges will have significant weight in the case of large polarons.
These approximationns are justified by the polaron amplitudes computed from first principles and shown in Figs. 2\textbf{f} and S6.
With this simplifications,
\begin{equation}
	\begin{array}{lll}{W}_{fi}({{{\bf{q}\nu}}};\omega ) &\propto&\mathop{\sum}\limits_{\beta =\pm 1}{\left\vert \, ({{{\bf{e}}}}\cdot [{{{{\bf{S}}}}}_{1,cv\nu }({{{\bf{k}_{\mathrm{VBM}}}}},{{{\bf{q}}}})+{{{{\bf{S}}}}}_{2,cv\nu \beta }({{{\bf{k}_{\mathrm{VBM}}}}},{{{\bf{q}}}})]) \, \frac{1}{N_p} \,\sum_{\mathbf{k}} A_{c\mathbf{k+q}}^{*} A_{v\mathbf{k}} \, \right\vert }^{2}\\ && \times \,\delta ({E }_{f}-{E }_{i}-\hbar \omega +\beta \hbar {\omega }_{{{{\bf{q}}}}\nu })~.\end{array}
\end{equation}
In the periodic case without polaron localization, $A_{v\mathbf{k}}=\delta_{\mathbf{k},\mathbf{k}_{\mathrm{VBM}}}$ and $A_{c\mathbf{k+q}}=\delta_{\mathbf{k+q},\mathbf{k}_{\mathrm{CBM}}}$,
and the standard expression is recovered:
\begin{equation}
	\begin{array}{lll}{W}^{0}_{fi}({{{\bf{q}\nu}}};\omega ) &\propto& \delta_{\mathbf{q}, \mathbf{k}_{\mathrm{CBM}}-\mathbf{k}_{\mathrm{VBM}}} \mathop{\sum}\limits_{\beta =\pm 1}{\left\vert \, ({{{\bf{e}}}}\cdot [{{{{\bf{S}}}}}_{1,cv\nu }({{{\bf{k}_{\mathrm{VBM}}}}},{{{\bf{q}}}})+{{{{\bf{S}}}}}_{2,cv\nu \beta }({{{\bf{k}_{\mathrm{VBM}}}}},{{{\bf{q}}}})]) \right\vert }^{2}\\ && \times \,\delta ({E }_{f}-{E }_{i}-\hbar \omega +\beta \hbar {\omega }_{{{{\bf{q}}}}\nu })~.\end{array}
\end{equation}
Thus, the relative change in the phonon-mediated recombination rate accross the indirect band gap at $\mathbf{q}_{\mathrm{gap}}=\mathbf{k}_{\mathrm{CBM}}-\mathbf{k}_{\mathrm{VBM}}$ due to polaron localization can be estimated by:
\begin{equation} \label{eq:recomb_final}
	\frac{{W}_{fi}({{{\bf{q}_{\mathrm{gap}}\nu}}})}{{W}^{0}_{\mathbf{k}+\mathbf{q}_{\mathrm{gap}}, \mathbf{k}}({{{\bf{q}_{\mathrm{gap}}\nu}}})}
=
{\left\vert \frac{1}{N_p} \,\sum_{\mathbf{k}} A_{m\mathbf{k+q}_{\mathrm{gap}}}^{*} A_{n\mathbf{k}} \, \right\vert }^{2} ~.
\end{equation}

A numerical evaluation of the overlap in Eq.~\ref{eq:recomb_final} yields $\frac{{W}_{fi}({{{\bf{q}_{\mathrm{gap}}\nu}}})}{{W}^{0}_{\mathbf{k}+\mathbf{q}_{\mathrm{gap}}, \mathbf{k}}({{{\bf{q}_{\mathrm{gap}}\nu}}})}=0.32$. This means that carrier localization upon polaron formation causes a threefold reduction of the electron-hole recombination rates in Cs$_2$AgBiBr$_6$.

\section*{Supplementary Note 7: Diffuse scattering from polaron distortions} \label{sec:supp7}

In order to simulate the polaronic contribution to the diffuse scattering intensity, we start from the recent manuscripts by Guzelturk and coworkers Ref.~\cite{Guzelturk2021} (Ref. 31 of the main text) and by de Cotret and coworkers Ref.~\cite{deCotret2022}, which represent the current state-of-the-art in this field. 
In these studies, simplified gaussian or exponential functions have been used to model the polaronic displacement pattern and interpret experimental data.
Here we go slightly beyond these efforts by using the \textit{ab initio} displacements obtained in our calculations.
In order to rationalize our findings, we compare our numerical data with simplified models, as we detail below.

The total scattering intensity by an extended crystal is given by:
\begin{equation} \label{eq:i0}
	I^{0}(\mathbf{Q}) = \left| \sum_{\kappa p} f_{\kappa}(\mathbf{Q}) \, e^{i\mathbf{Q}\cdot \left( \mathbf{R}_{p} + \boldsymbol{\tau}^{0}_{\kappa} \right)}\right|^2 ~,
\end{equation}
where $\mathbf{Q}$ is the scattering vector .
The summation runs over all atoms $\kappa$ in the unit cell and over all $p$ unit cells with lattice vectors $\mathbf{R}_{p}$ in the crystal. 
The atomic form factors are given by $f_{\kappa}(\mathbf{Q})$.
The scattering intensity in the distorted polaronic configuration is given by:
\begin{equation} \label{eq:i}
	I(\mathbf{Q}) = \left| \sum_{\kappa p} f_{\kappa}(\mathbf{Q}) \, e^{i\mathbf{Q}\cdot \left( \mathbf{R}_{p} + \boldsymbol{\tau}^{0}_{\kappa} + \Delta \boldsymbol{\tau}_{\kappa p}\right)}\right|^2 ~.
\end{equation}
For small displacements, the exponential factor in Eq.~\ref{eq:i} can be expanded as:
\begin{equation}
	e^{i\mathbf{Q}\cdot(\mathbf{R}_{p}+\boldsymbol{\tau}^{0}_{\kappa}+\Delta\boldsymbol{\tau}_{\kappa p})} = (1 + i\mathbf{Q}\cdot \Delta \boldsymbol{\tau}_{\kappa p} ) 
	\, e^{i\mathbf{Q}\cdot(\mathbf{R}_{p}+\boldsymbol{\tau}^0_{\kappa})} + \mathcal{O}(\Delta \boldsymbol{\tau}^2) ~,
\end{equation}
so that at linear order in the displacements, the scattering intensity reads:
\begin{equation} \label{eq:linear_tau}
	I(\mathbf{Q}) = \left| \sum_{p\kappa} f_{\kappa}(\mathbf{Q}) \, e^{i\mathbf{Q}\cdot(\mathbf{R}_{p}+\boldsymbol{\tau}^{0}_{\kappa})} + i \sum_{p\kappa} \mathbf{Q}\cdot \Delta \boldsymbol{\tau}_{\kappa p} \, f_{\kappa}(\mathbf{Q}) \, e^{i\mathbf{Q}\cdot(\mathbf{R}_{p}+\boldsymbol{\tau}^{0}_{\kappa})} \right|^2 = \left| F^{0}(\mathbf{Q}) + F^{\mathrm{P}}(\mathbf{Q}) \right|^2 ~.
\end{equation}

The total scattering vector can be expressed as:
\begin{equation}
	\mathbf{Q} = \mathbf{G}+\mathbf{q} ~,
\end{equation}
where $\mathbf{G}$ is a reciprocal lattice vector and $\mathbf{q}$ belongs to the first Brillouin zone.
Close to the $\mathbf{G}\neq 0$ vectors (Bragg peaks), the scalar product can be approximated as:
\begin{equation}
	\mathbf{Q} \cdot \Delta \boldsymbol{\tau}_{\kappa p} = (\mathbf{G}+\mathbf{q}) \cdot \Delta \boldsymbol{\tau}_{\kappa p} \approx \mathbf{G} \cdot \Delta \boldsymbol{\tau}_{\kappa p} ~,
\end{equation}
so that $F^{\mathrm{P}}(\mathbf{Q})$ simplifies to:
\begin{equation}
	F^{\mathrm{P}}(\mathbf{Q}) = i \sum_{p\kappa} \mathbf{G}\cdot \Delta \boldsymbol{\tau}_{\kappa p} \, f_{\kappa}(\mathbf{G}+\mathbf{q}) \, e^{i(\mathbf{G}+\mathbf{q})\cdot(\mathbf{R}_{p}+\boldsymbol{\tau}^{0}_{\kappa})} ~.
\end{equation}

Assuming there is a single atomic species per unit cell,
\begin{equation} \label{eq:discrete_fourier}
	F^{\mathrm{P}}(\mathbf{Q}) = i f(\mathbf{G}+\mathbf{q}) \, \mathbf{G}\cdot \sum_{p} \Delta \boldsymbol{\tau}_{p} \, e^{i\mathbf{q} \cdot \mathbf{R}_{p}} ~,
\end{equation}
and in the continuous limit ($\Delta\boldsymbol{\tau}_{p}\rightarrow \mathbf{u}(\mathbf{r})$, $\sum_{p}\rightarrow\int d\mathbf{r}$) relevant for long-range polaronic distortions:
\begin{equation}
	F^{\mathrm{P}}(\mathbf{Q}) = i f(\mathbf{G}+\mathbf{q}) \, \mathbf{G} \cdot \int d\mathbf{r} \, \mathbf{u}(\mathbf{r}) e^{i\mathbf{q} \cdot \mathbf{r}} = i f(\mathbf{G}+\mathbf{q}) \, \mathbf{G} \cdot \mathbf{u}(\mathbf{q}) ~,
\end{equation}
where $\mathbf{u}(\mathbf{q})$ is the Fourier transform of the displacement field.
For an inversion-symmetric displacement field, the cross-term in Eq.~\ref{eq:linear_tau} vanishes and the total scattering intensity is given by:
\begin{equation}
	I(\mathbf{Q}) = 
	\left| F^{0}(\mathbf{Q}) \right|^2 + \left| f(\mathbf{G}+\mathbf{q}) \, \mathbf{G} \cdot \mathbf{u}(\mathbf{q})  \right|^2 ~,
\end{equation}
so that we can get the difference in scattering intensity induced by polarons by substracting the scattering intensity of the original undistorted crystal:
\begin{equation} \label{eq:huang}
	\Delta I(\mathbf{Q}) = I(\mathbf{Q}) - I^0(\mathbf{Q}) = \left| f(\mathbf{G}+\mathbf{q}) \, \mathbf{G} \cdot \mathbf{u}(\mathbf{q})  \right|^2 ~.
\end{equation}
For ease of visualization, we follow Refs.~\cite{Guzelturk2021} and \cite{deCotret2022} and plot the normalized differential intensity, assuming that the scattering intensity without polaronic distortion falls off as $1/|\mathbf{q}|$:
\begin{equation} \label{eq:huangplot}
	\frac{\Delta I(\mathbf{Q})}{I^0(\mathbf{Q})} \propto |\mathbf{q}| \, \left| \mathbf{G} \cdot \mathbf{u}(\mathbf{q})  \right|^2.
\end{equation}

The simplified expression in Eq.~\ref{eq:huang} is equivalent to the so-called Huang diffuse scattering intensity well known in the context of long-range distortions associated with point defects [see e.g. Eq.~(11) of Ref.~\cite{Peisl1976}].
Interestingly, this expression shows that, depending on the shape of the displacement field, the scalar product will result in regions of reciprocal space where the differential intensity vanishes. This observation has been used in the context of point defects to study in detail the symmetries of the distortions [see e.g. Ref.~\cite{Dederichs1973}].
In the simplest case of a radial displacement field, Eq.~\ref{eq:huang} gives a dipole-like shape with vanishing $\Delta I(\mathbf{Q})$ for the $\mathbf{Q}\perp \mathbf{G}$ plane (cf. Fig.~3 of \cite{Peisl1976}).
Most importantly, 
in the case of a helical Bloch point, the in-plane rotation of the displacement field in real space directly results in a rotation of the reciprocal-space plane where the Huang intensity vanishes. The rotation of the plane is exactly given by $\gamma$, the helicity of the Bloch point. 

We illustrate this point in Supplementary Fig.~\ref{fig10}.
In the upper left panel, we show a cross section of a model radial displacement pattern with a gaussian decay,
and in the lower left panel we show a cross section of the corresponding normalized differential intensity obtained from Eq.~\ref{eq:huangplot}, close to a Bragg point along the $(0,1,0)$ direction. The dipole-like shape is clearly observed with a vanishing Huang scattering for $\mathbf{Q}\perp\mathbf{G}$.
In the upper middle panel, we show a cross section for a model Bloch point displacement field with a gaussian decay, in which the parameters have been set to mimic the large electron polaron in Cs$_2$AgBiBr$_6$ (see Fig.~S4). In particular, the helicity has been set to $\gamma=3\pi/4$. In the lower middle panel we show the corresponding normalized differential intensity, where the rotation of the dipole-like shape and the reciprocal-space plane where the Huang scattering vanishes is clearly observed.
In the upper right panel we show a cross section for the \textit{ab initio} displacement pattern corresponding to the large electron polaron (cf. Fig.~1\textbf{h}-\textbf{l} of the main text). In the lower right panel, we show the normalized differential intensity, where the Fourier transform in Eq.~\ref{eq:discrete_fourier} has been taken numerically from the \textit{ab initio displacements}.
The anisotropy and slight deviation from the perfect Bloch point displacements adds more complexity to the Huang scattering pattern, but the dipole-like shape and the rotation given by the helicity are clearly discernible.
This analysis shows that the anisotropy of the Huang scattering provides an unambiguous fingerprint of the finite helicity and represents a direct physical implication of the topological nature of the polarons, which could serve as a direct experimental verification of our results.


\clearpage

\begin{figure*}
  \centering
  \includegraphics[width=0.8\textwidth]{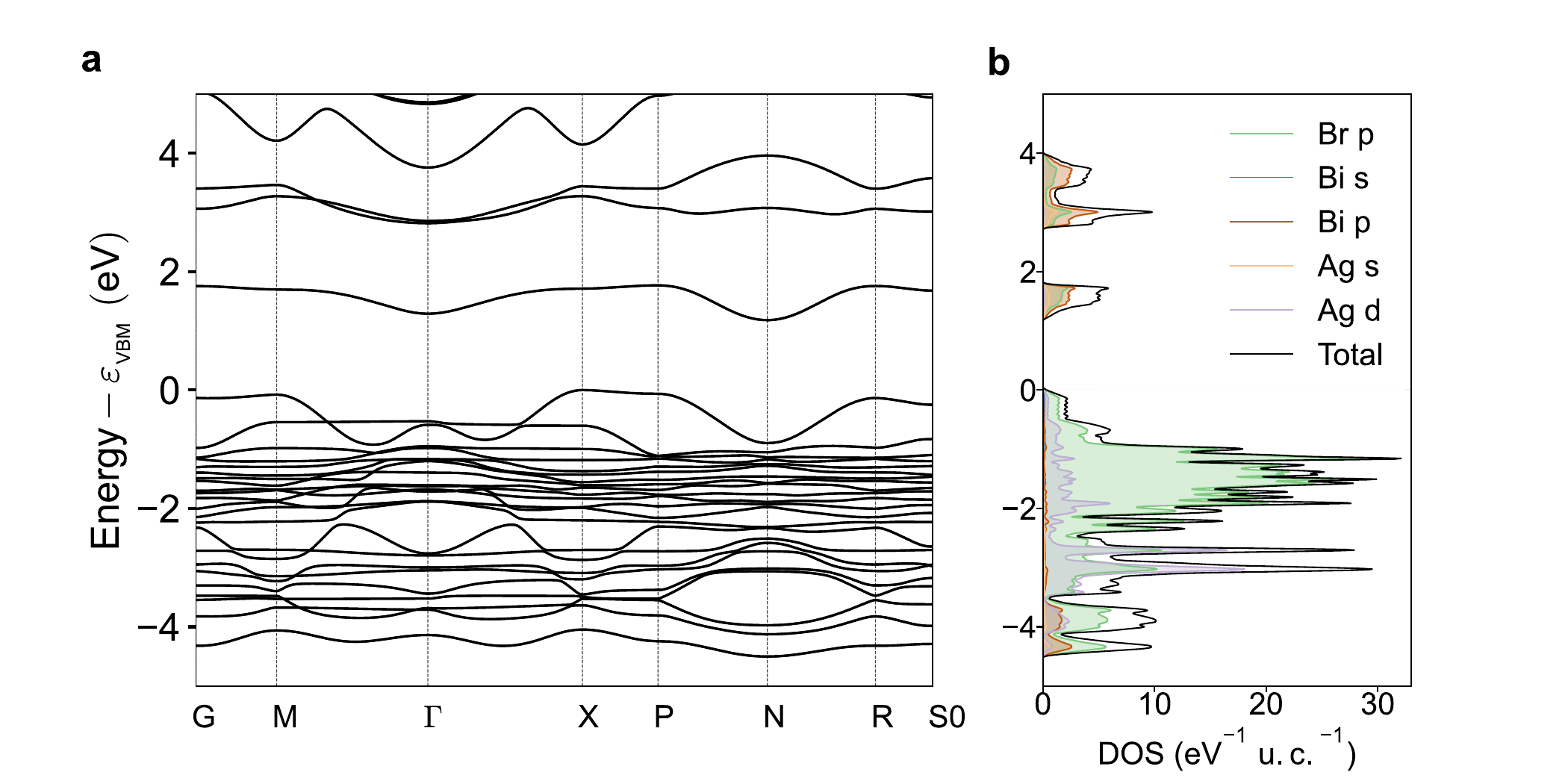}
  \caption{\textbf{Electronic structure of Cs$_2$AgBiBr$_6$}. \textbf{a} DFT band structure of Cs$_2$AgBiBr$_6$. \textbf{b} Projected density of states for the band structure shown in (a). 
  \label{fig:bands_dos}}
\end{figure*}

\clearpage

\begin{figure*}
  \centering
  \includegraphics[width=1.0\textwidth]{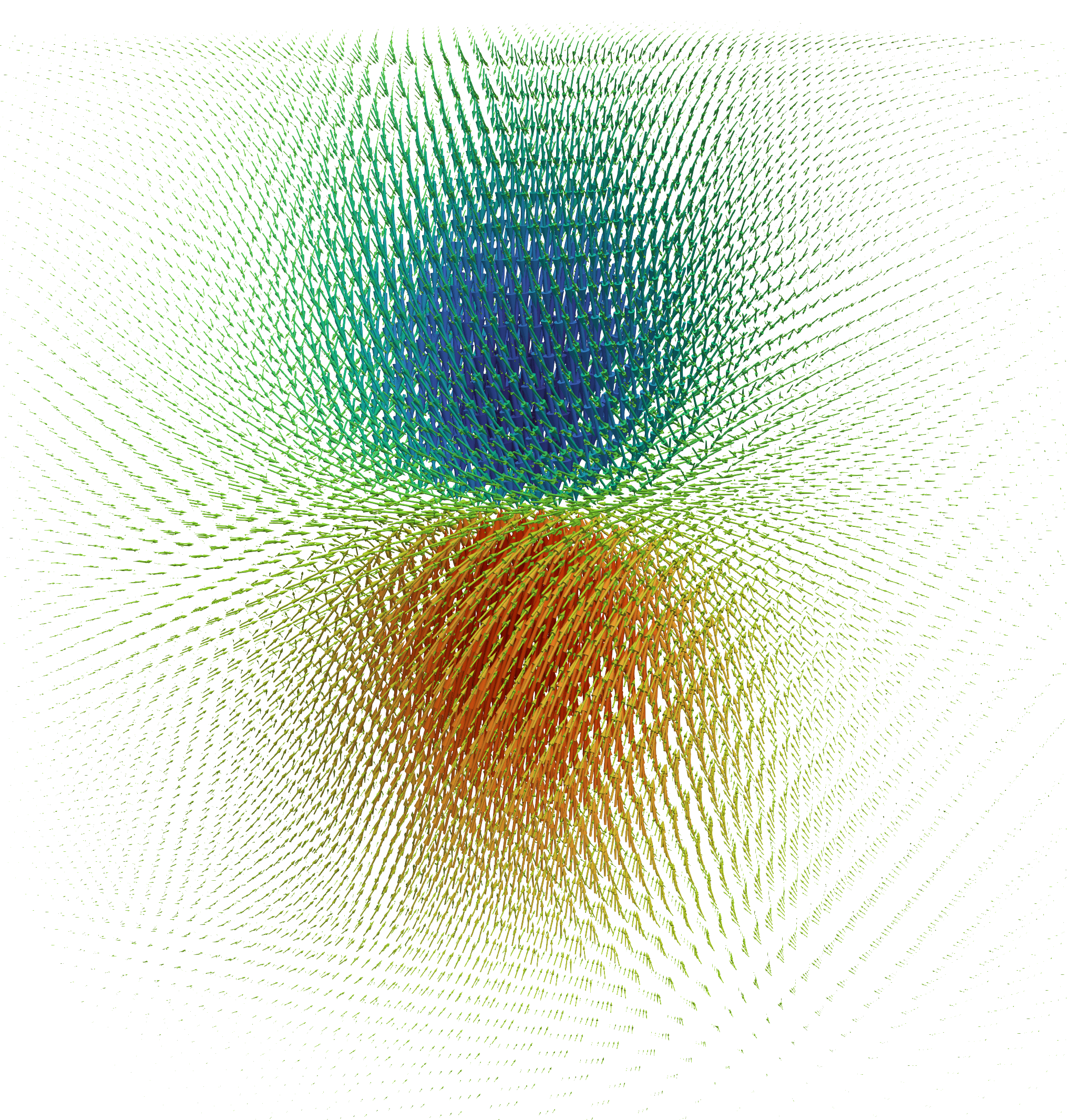}
  \caption{\textbf{Enlarged view of the displacement pattern associated with the large electron polaron in Cs$_2$AgBiBr$_6$, shown in Fig.~1h of the main text}. The plot represents the displacements of the 32,000 Ag atoms within a 20$\times$20$\times$20 pseudocubic supercell. 
  \label{fig:enlarged_electron}}
\end{figure*}

\clearpage

\begin{figure*}
  \centering
  \includegraphics[width=1.0\textwidth]{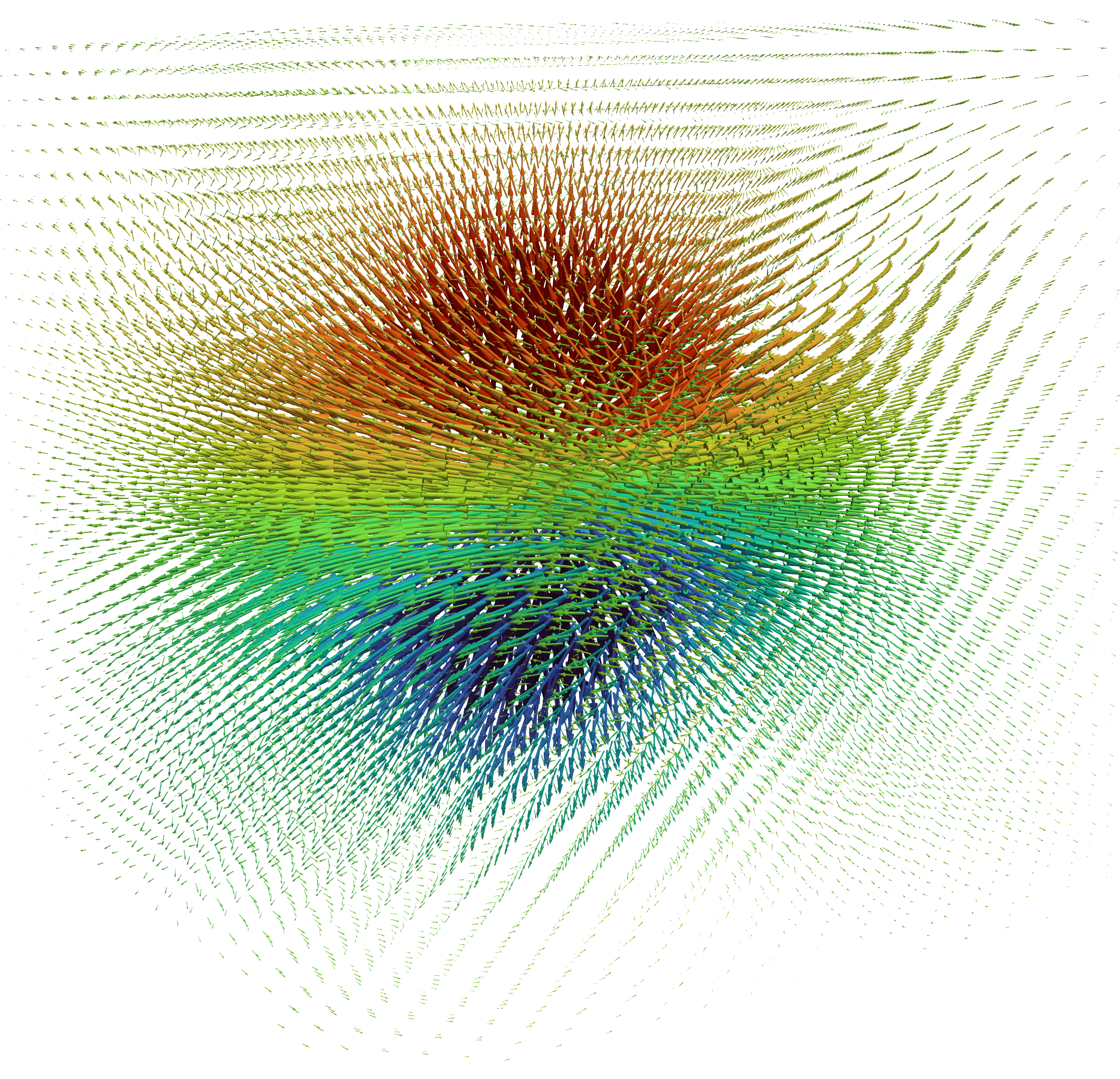}
  \caption{\textbf{Enlarged view of the displacement pattern associated with the large hole polaron in Cs$_2$AgBiBr$_6$, shown in Fig.~1m of the main text}. The plot represents the displacements of the 23,328 Ag atoms within a 18$\times$18$\times$18 pseudocubic supercell.
  \label{fig:enlarged_hole}}
\end{figure*}

\clearpage

\begin{figure*}
  \centering
  \includegraphics[width=1.0\textwidth]{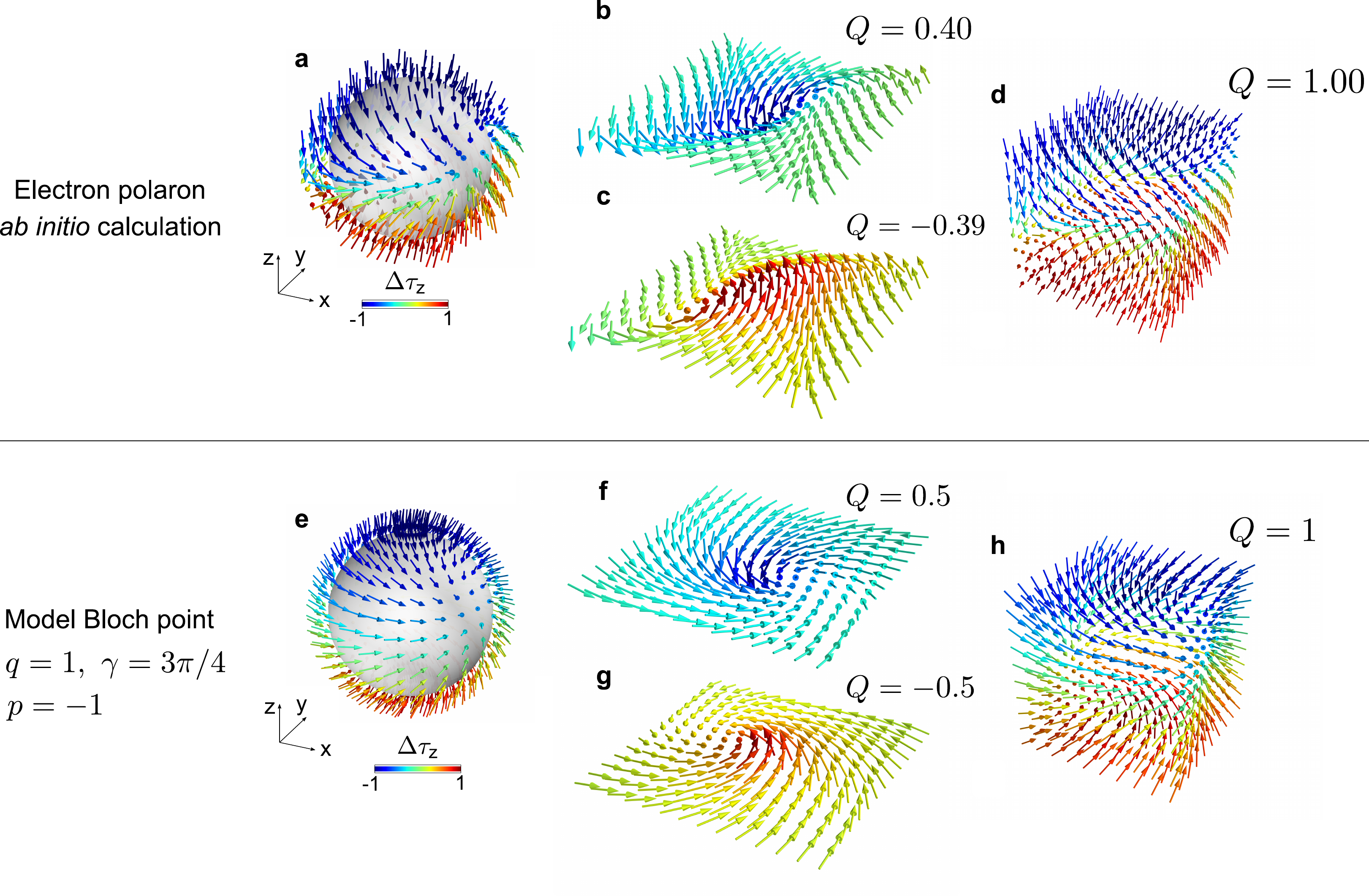}
  \caption{\textbf{Topology of large electron polaron in Cs$_2$AgBiBr$_6$ and helical Bloch point.}
  \textbf{a} Normalized atomic displacements ($\Delta\hat{\boldsymbol{\tau}} = \Delta \boldsymbol{\tau}/|\Delta \boldsymbol{\tau}|$) associated with the large electron polaron in Cs$_2$AgBiBr$_6$ shown in Fig.~\textbf{1h} of the main text, restricted to a spherical surface around the polaron center. The radius of this surface is 40~\AA, and only displacements of the Ag atoms are shown for clarity.  \textbf{b}, \textbf{c} Cross-sections of the displacements along the planes $z=-5.8$~\AA\ and $z=5.8$~\AA, respectively. The 2D topological charges computed for these cross-sections are indicated in each panel. Note that, in the case of \textit{ab initio} calculations, the deviations from the analytical (integer or half-integer) values of the topological charge are due to the finite size of the supercell employed in the numerical integration. A numerical test of this effect is shown in Supplementary Fig.~S11. The charge is evaluated by performing a cubic interpolation of the displacement field, and by carrying out the integration over a square with size 60~\AA. \textbf{d} 3D rendering of the normalized Ag displacements in a cube of size 40~\AA\ enclosing the polaron center. The 3D topological charge computed over the surface of this cube is indicated.  \textbf{e}, \textbf{f}, \textbf{g}, \textbf{h} Same as \textbf{a}-\textbf{d} but for a model Bloch point.  
  The Bloch points is described using the expression of Refs.~\citenum{Doring1968,Pylypovskyi2012}: $\mathbf{M}/|\mathbf{M}| = \mathbf{m} = (\sin \Theta(\mathbf{r}) \cos \Phi(\mathbf{r}), \sin \Theta(\mathbf{r}) \sin \Phi(\mathbf{r}), \cos \Theta(\mathbf{r}))$ and $ \Theta(\theta) = p\theta + \pi (1-p)/2 ~, \Phi(\phi) = q\phi + \gamma ~, $ where $p$ is the polarity of the Bloch point, $q$ is the vorticity, $\gamma$ is the helicity. These parameters are related to the topological charge via $Q=-pq$. The polarity and vorticity of the \textit{ab initio} Bloch point is obtained by averaging $d\Theta/d\theta$ and $d\Phi/d\phi$ over a longitudinal path between the poles ($\phi=0$ and $\theta$ varying from $0$ to $\pi$) and over an equatorial path ($\theta=0$ and $\phi$ varying from $0$ to $2\pi$), respectively. The helicity is obtained as $\gamma = (2\pi)^{-1}\int_0^{2\pi} [\Phi(\phi)-q\phi] \, d\phi$. For the electron polaron in \textbf{a} the results are $p=-1$, $q=1$ and $\gamma=3\pi/4$. These parameters have been used to represent the model Bloch point in \textbf{e}-\textbf{h}.}
  \label{fig:bloch_electron}
\end{figure*}

\clearpage

\begin{figure*}
  \centering
  \includegraphics[width=1.0\textwidth]{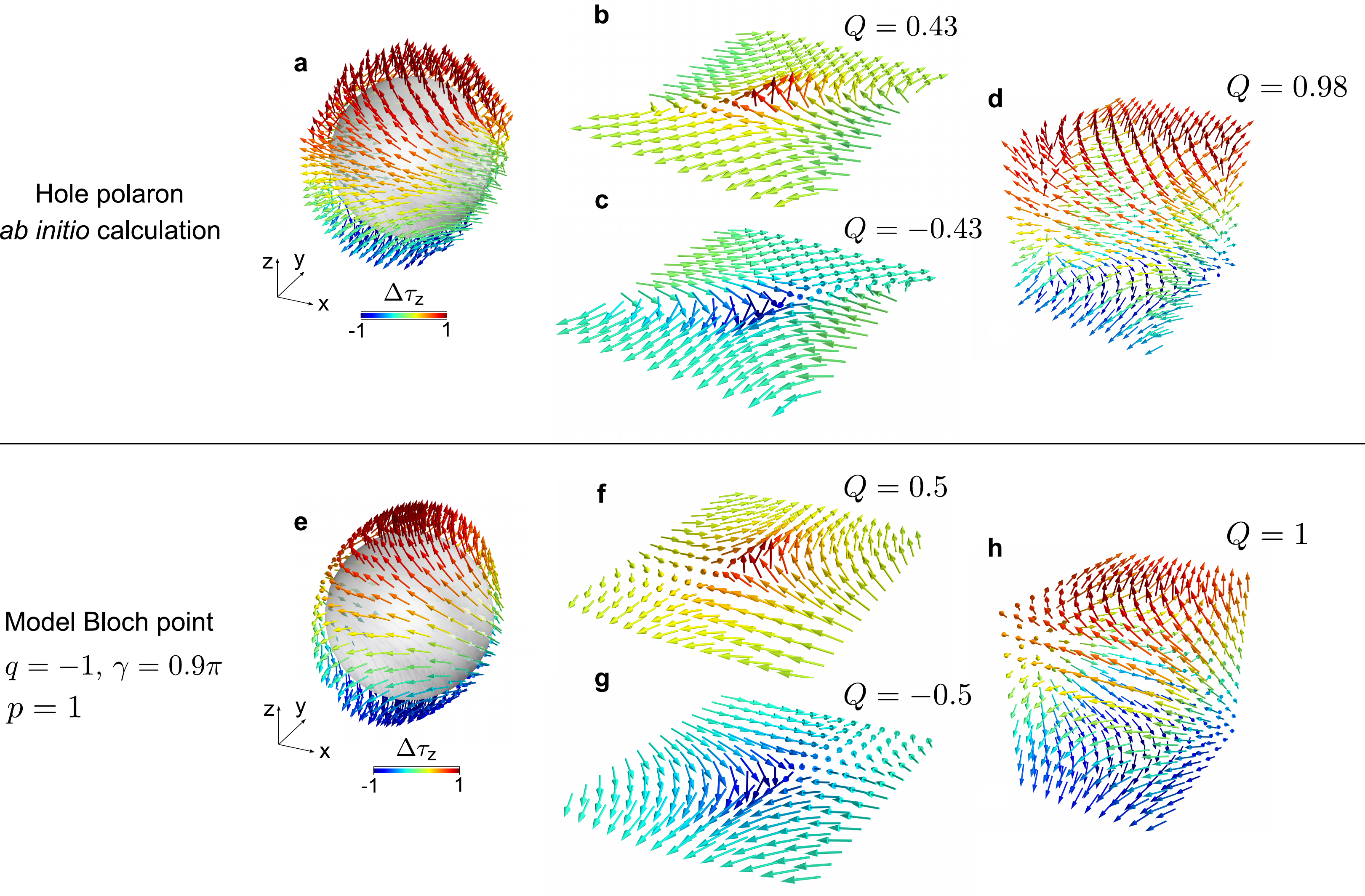}
  \caption{\textbf{Topology of large hole polaron in Cs$_2$AgBiBr$_6$ and helical Bloch point.}
   Same as Fig.~\ref{fig:bloch_electron}, but for the large hole polaron shown in Fig.~\textbf{1m} of the main text. In this case, we obtain a polarity $p=1$, a vorticity $q=-1$ and a helicity $\gamma=0.9\pi$.
   \label{fig:bloch_hole}}
\end{figure*}

\clearpage

\begin{figure*}
  \centering
  \includegraphics[width=1.0\textwidth]{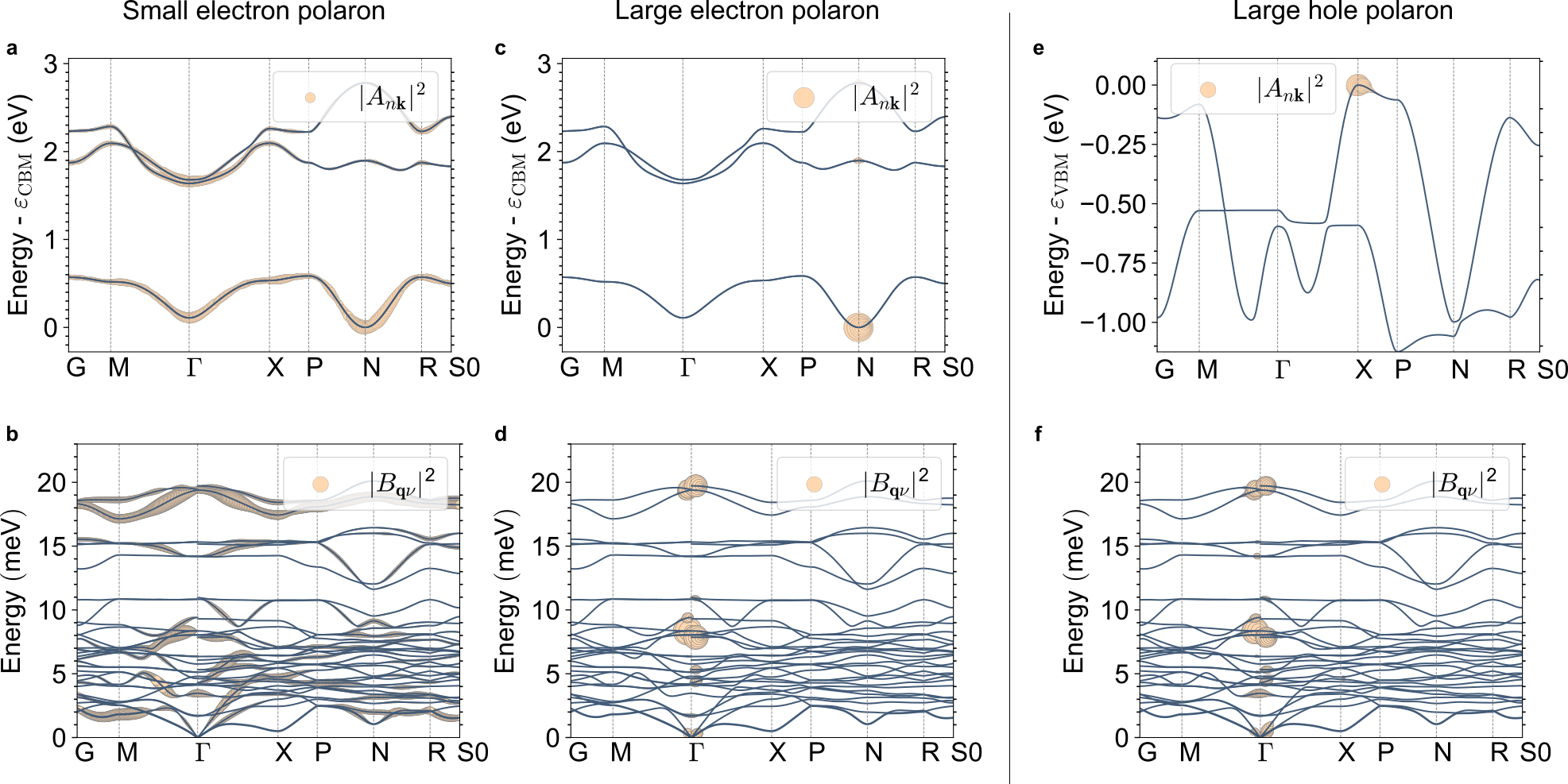}
  \caption{\textbf{Spectral decomposition of electron and hole polarons in Cs$_2$AgBiBr$_6$.} \textbf{a} Expansion coefficients $|A_{n{\bf k}}|^2$ of the wavefunction of the small electron polaron shown in Fig.~\textbf{2} of the main text. These coefficients are superimposed to the electronic band structure, and the radius of each circle is proportional to $|A_{n{\bf k}}|^2$. \textbf{b} Expansion coefficients $|B_{{\bf q}\nu}|^2$ of the atomic displacements associated with the small electron polaron in Fig.~\textbf{2} of the main text. The coefficients are superimposed to the phonon band structure. \textbf{c}, \textbf{d} and \textbf{e}, \textbf{f} Same as \textbf{a}, \textbf{b} but for the large electron polaron, and for the large hole polaron in Fig.~\textbf{1} of the main text, respectively. 
  \label{fig:Ank_Bqv}}
\end{figure*}

\clearpage

\begin{figure*}
  \centering
  \includegraphics[width=1.0\textwidth]{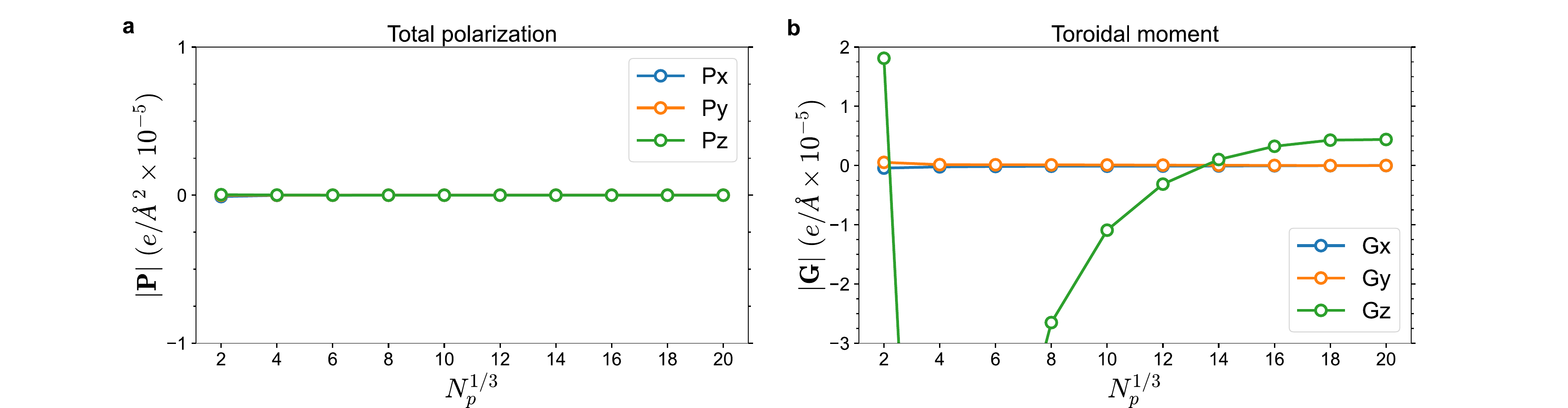}
  \caption{\textbf{Toroidal moment of the small electron polaron in Cs$_2$AgBiBr$_6$}. \textbf{a} Total dielectric polarization associated with the small electron polaron in Cs$_2$AgBiBr$_6$, as a function of supercell size. For each atom, the dipole moment is given by $P_{\kappa\alpha}=\sum_{\beta} Z^{*}_{\kappa \alpha\beta}\Delta\tau_{\kappa p\beta}$, where $Z^{*}_{\kappa\alpha\beta}$ represents the Born effective charge tensor. The total polarization is then obtained as $\mathbf{P} = \sum_{\kappa p}  \mathbf{P}_{\kappa p}/(N_p\Omega)$, where $\Omega$ is the volume of the unit cell, and $N_p$ is the number of unit cells in the supercell. This panel shows that the macroscopic polarization associated with the small electron polaron vanishes. \textbf{b} Ferroelectric toroidal moment associated with the small electron polaron in Cs$_2$AgBiBr$_6$.
  To evaluate this quantity, we employ the definition of toroidal moment of Ref.~\citenum{Naumov2004} adapted to the polaron case \cite{Shimada2020}: $\mathbf{G} = (2N_p\Omega)^{-1} \sum_{\kappa p} \mathbf{r}_{\kappa p} \times \mathbf{P}_{\kappa p}$, where $\mathbf{r}_{\kappa p}$ is the position of the atom $\kappa, p$ with respect to the Bi atom at the center of the polaron. The panel shows that the in-plane components of the ferrotoroidic moment vanish, while the out-of-plane component converges to a finite number. We note that very large supercells are needed to reach a converged result, in this case a supercell consisting of $20\times 20\times 20$ unit cells. 
 \label{fig:toroidal}}
\end{figure*}

\clearpage

\begin{figure*}
  \centering
  \includegraphics[width=0.75\textwidth]{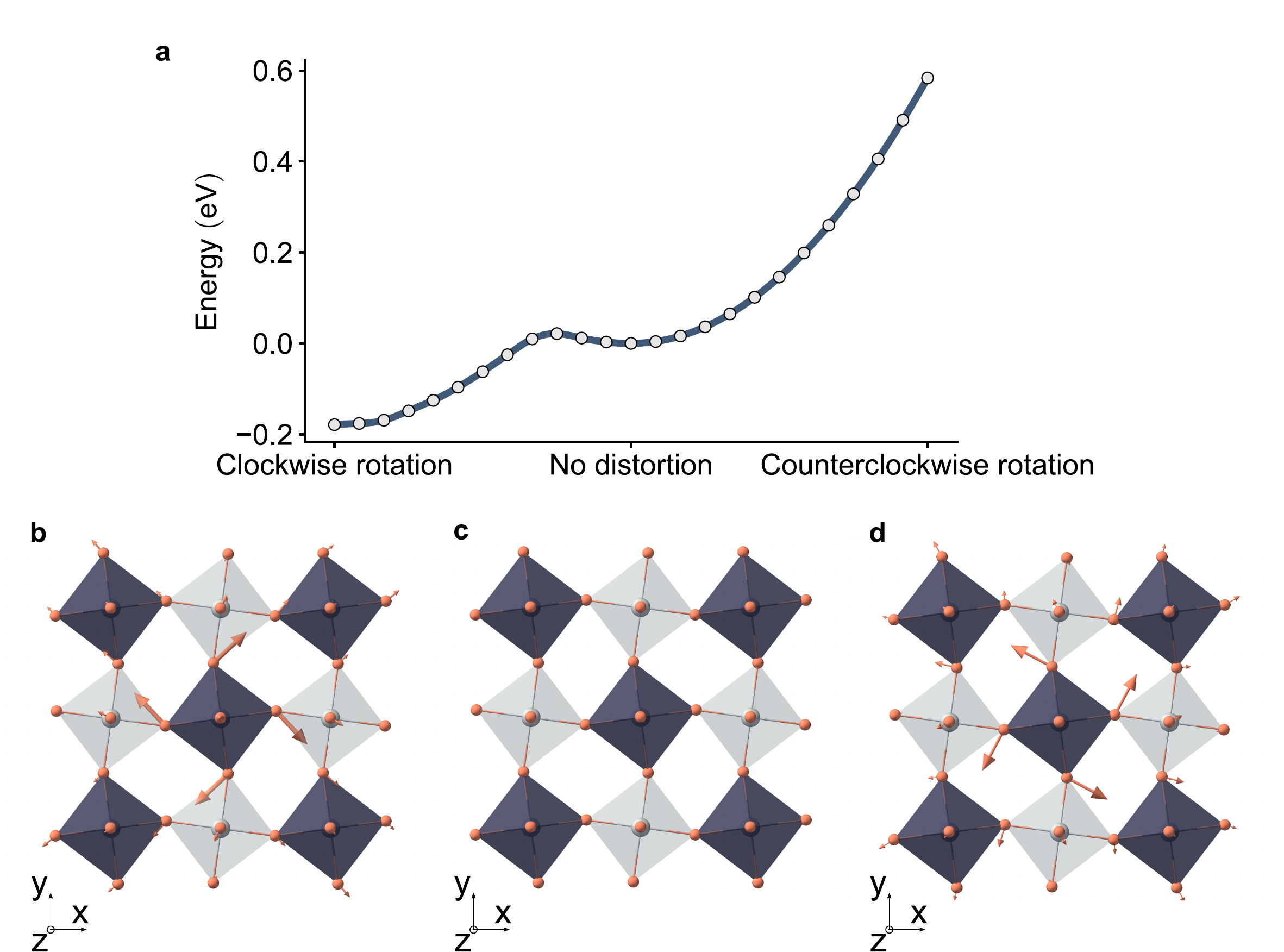}
  \caption{\textbf{Rotation asymmetry of the small electron polaron in Cs$_2$AgBiBr$_6$.} BiBr$_6$ octahedra in the Cs$_2$AgBiBr$_6$ structure are all oriented in the same way (cf. Fig.~1\textbf{a} of the main text), and the small electron polaron localizes around one Bi atom (cf. Fig~2 of the main text). This figure shows that the octahedral rotations associated with small electron polarons always point in the same direction. \textbf{a} Polaron energy landscape along a configuration space path starting from the small electron polaron solution (\textbf{b}), passing through the undistorted configuration (\textbf{c}), and ending with a hypothetical configuration in which the displacements are localized around the same Bi atom but the rotation direction of the Br octahedron has been inverted (\textbf{d}).  This mirror configuration is more energetic that the undistorted configuration and the original polaron solution, and thus is not stable. This implies that the sense of octahedral rotation associated with each polaron, and hence the sign of the toroidal moment discussed in Fig.~\textbf{S5}, does not change upon e.g. hopping to adjacent sites or dense-packing of many polarons.}
\end{figure*}

\clearpage

\begin{figure*}
	\centering
	\includegraphics[width=0.6\textwidth]{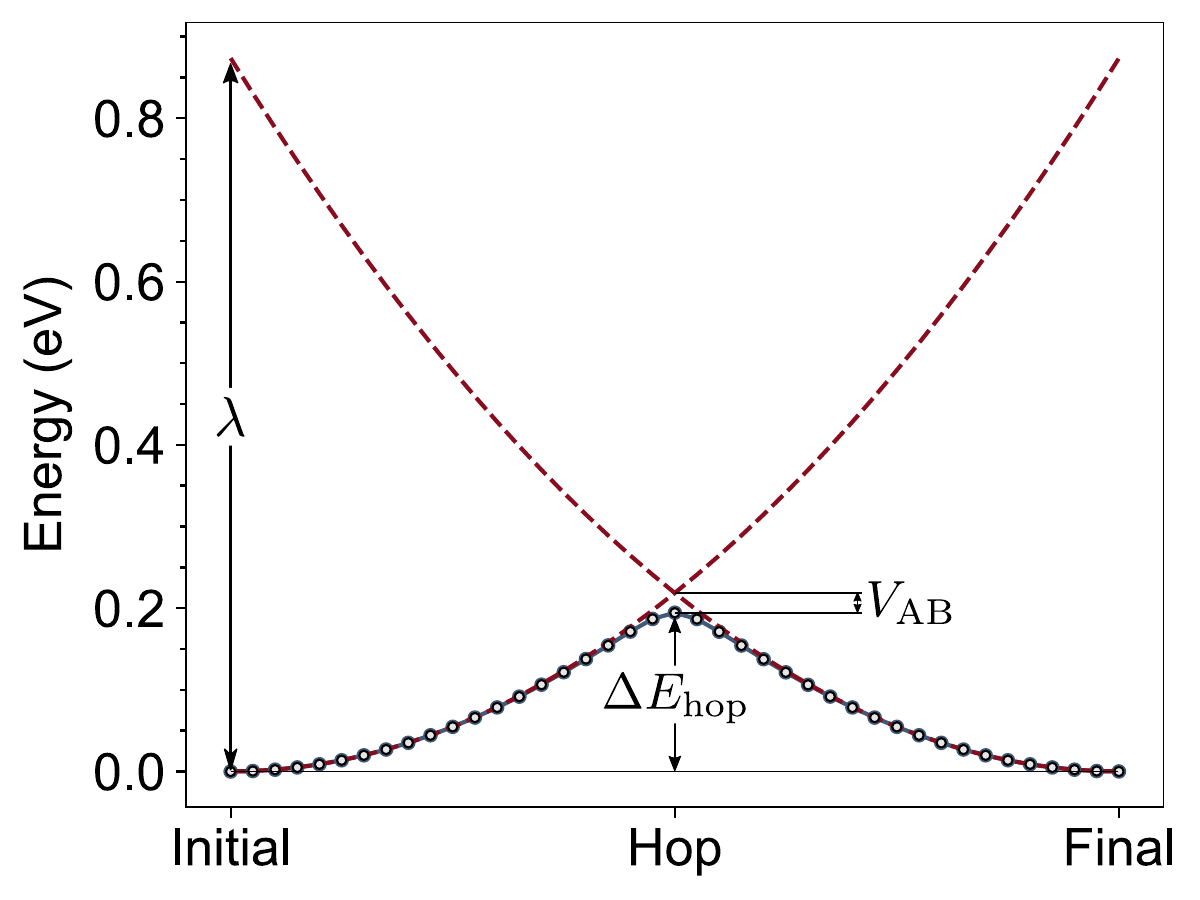}
	\caption{\textbf{Parabolic fit of the potential energy surface minima.} The potential energy surface is the same as in Fig.~2\textbf{i} of the main text. The electronic coupling energy $V_{\mathrm{AB}}$, the reorganization energy $\lambda$, and the adiabatic hopping energy barrier $\Delta E_{\mathrm{hop}}$ are indicated.}
	\label{fig9}
\end{figure*}

\clearpage

\begin{figure*}
	\centering
	\includegraphics[width=1.0\textwidth]{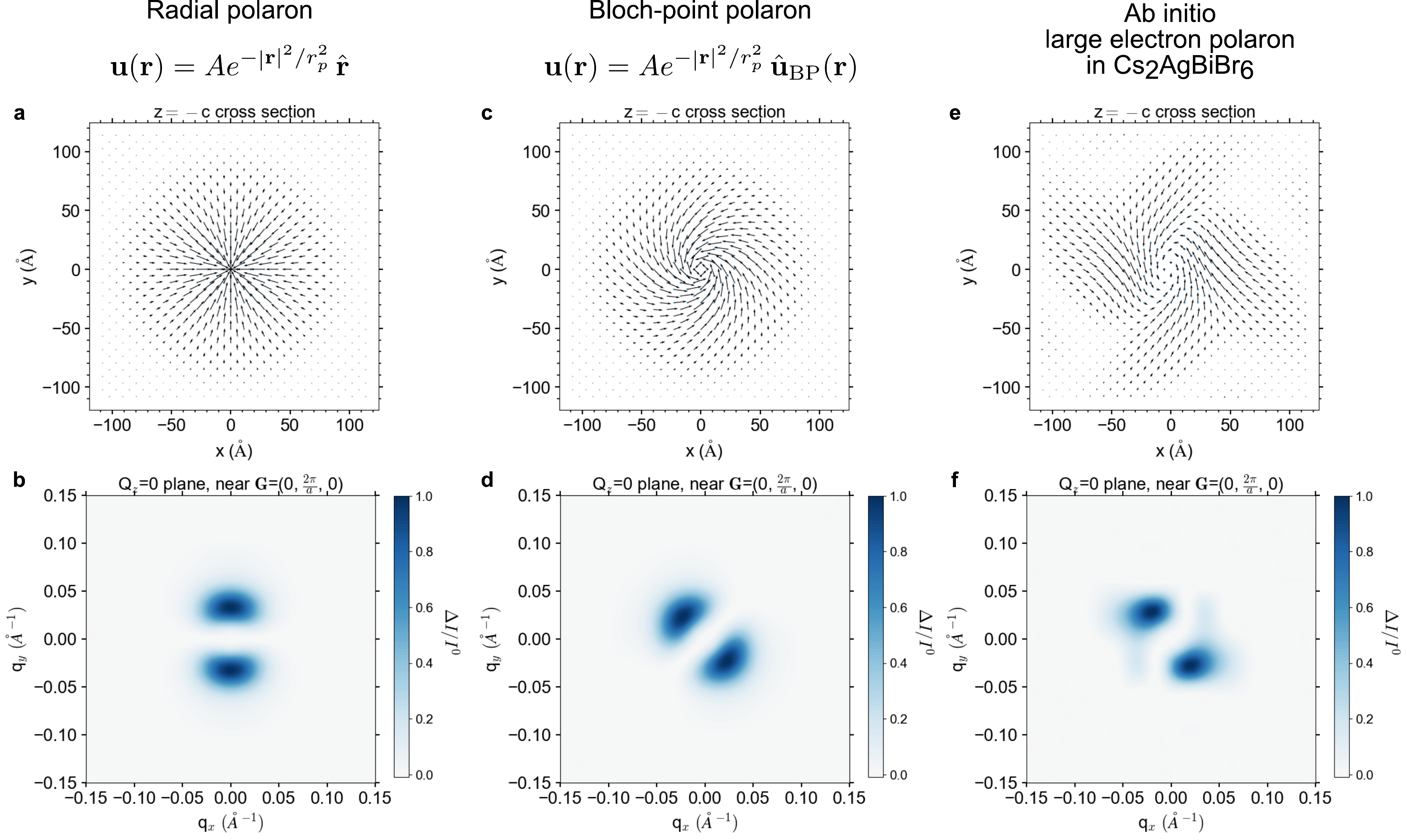}
	\caption{\textbf{Huang scattering of topological polarons with finite helicity.} \textbf{a} Cross section of the real space displacements corresponding to a radial displacement pattern with a gaussian decay. The parameters have been set to $A=0.001$ \AA~ and $r_p = 63$ \AA. \textbf{b} Cross section of the corresponding normalized differential intensity obtained from Eq.~\ref{eq:huangplot}, close to a Bragg point along the $(0,1,0)$ direction. \textbf{c} and \textbf{d} Same as \textbf{a} and \textbf{b} but for a helical Bloch point displacement pattern with a gaussian decay. $\hat{\mathbf{u}}_{\mathrm{BP}}$ represents the unit vector of the Bloch-point field given in Fig.~S4. The same parameters as in Fig.~S4 have been set, namely $p=-1$, $q=1$ and $\gamma=3\pi/4$. \textbf{c} and \textbf{d} Same as \textbf{a} and \textbf{b} but for the numerical displacement pattern obtained from \textit{ab initio} calculations for the large electron polaron in Cs$_2$AgBiBr$_6$ shown in Fig.~1\textbf{h}--\textbf{l} of the main text. For simplicity, only the displacements of the Ag atoms have been considered.}
	\label{fig10}
\end{figure*}

\clearpage

\begin{figure*}
	\centering
	\includegraphics[width=0.8\textwidth]{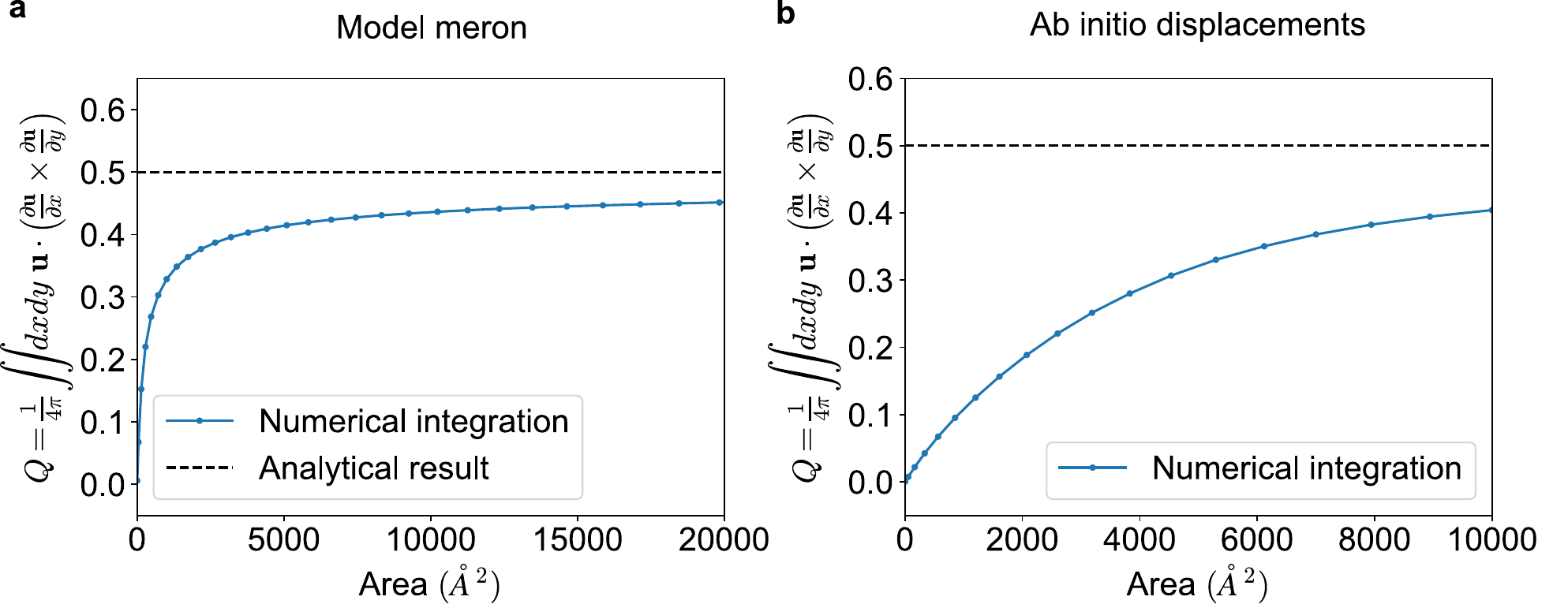}
	\caption{\textbf{Numerical convergence of topological charge.} \textbf{a} Numerical integration of the topological charge corresponding to the model meron shown in Fig.~S4\textbf{f}, as a function of the integration area. The analytical result is represented by the dashed line. \textbf{b} Numerical integration of the topological charge for the \textit{ab initio} displacement field over the two-dimensional cross-section of the large electron polaron shown in Fig.~S4\textbf{b}. The integration area is limited by the supercell size in the calculation.}
	\label{fig1}
\end{figure*}

\clearpage

\bibliography{bibliography}

\end{document}